\documentclass[twocolumn,showpacs,prb,aps,epsf,amsmath,amssymb]{revtex4}
\usepackage{graphicx}
\usepackage{amsbsy}
\newlength{\onewidth}
\usepackage{tensor}
\usepackage{multirow}
\usepackage{epsfig}
\begin{document}

\title{Valley degeneracy in biaxially strained aluminum arsenide quantum wells}

\author{S.~Prabhu-Gaunkar$^1$}
\author{S.~Birner$^2$}
\author{S.~Dasgupta$^2$}
\author{C.~Knaak$^2$}

\author{M. Grayson$^{1}$}
\affiliation{
$^1$Electrical Engineering and Computer Science, Northwestern University, Evanston, IL 60208 USA\\
$^2$Walter Schottky Institut, Technische Universit\"at M\"unchen, Garching, D-85748 Germany
}

\begin{abstract}
This paper describes a complete analytical formalism for calculating electron subband energy and degeneracy in strained multi-valley quantum wells grown along any orientation with explicit results for AlAs quantum wells. In analogy to the spin index, the valley degree of freedom is justified as a pseudospin index due to the vanishing intervalley exchange integral. A standardized coordinate transformation matrix is defined to transform between the conventional-cubic-cell basis and the quantum well transport basis whereby effective mass tensors, valley vectors, strain matrices, anisotropic strain ratios, piezoelectric fields, and scattering vectors are all defined in their respective bases. The specific cases of (001)-, (110)-, and (111)-oriented aluminum arsenide (AlAs) quantum wells are examined, as is the unconventional (411) facet, which is of particular importance in AlAs literature.  Calculations of electron confinement and strain in the (001), (110), and (411) facets determine the critical well width for crossover from double- to single-valley degeneracy in each system. The biaxial Poisson ratio is calculated for the high-symmetry lower Miller index (001)-, (110)-, and (111)-oriented QWs. An additional shear strain component arises in the higher Miller index (411)-oriented QWs and we define and solve for a shear-to-biaxial strain ratio. The notation is generalized to address non-Miller-indexed planes so that miscut substrates can also be treated, and the treatment can be adapted to other multi-valley biaxially strained systems. To help classify anisotropic intervalley scattering, a valley scattering primitive unit cell is defined in momentum space which allows one to distinguish purely in-plane momentum scattering events from those that require an out-of-plane momentum component.

\end{abstract}


\maketitle

\section{Introduction}

\begin{figure}[b]
\includegraphics[width=\columnwidth]{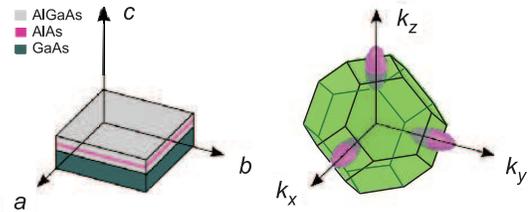}
	\caption{ (Left) Real space depiction of AlAs QW structures defining the transport basis axes, $a, b, c$.  (Right) Brillouin zone for bulk AlAs defining the conventional-cubic-cell basis $x, y, z$. Note the three degenerate valleys that are occupied with electrons as indicated by ellipsoidal equi-energy contours.}
	\label{fig:brill}
\end{figure}

Calculating valley degeneracy in a quantum well requires a comprehensive treatment of strain, quantum confinement, and piezoelectric fields since all contribute at comparable energy scales. Of the various multi-valley semiconductors, the indirect-bandgap zinc blende semiconductor aluminum arsenide (AlAs) with its bulk threefold valley degeneracy is of particular interest 
(Fig.~\ref{fig:brill}) because its heavy anisotropic electron mass allows for large interaction effects \cite{shayeganreview}, and its near perfect lattice match to GaAs substrates allows for high-mobility, modulation-doped quantum wells (QWs)  \cite{smithprb,lay}. AlAs/AlGaAs QWs can reach mobilities of the order of $\mu$ = 100,000 cm$^{2}$/Vs in (001)-facet QWs \cite{dasgupta} and in the high mobility direction of anisotropic (110)-facet QWs \cite{dasgupta2}. Unconventional facets such as (411) have also proven useful in identifying exchange effects like quantum Hall ferromagnetism in AlAs QWs \cite{depoortereScience2000,vaki2}.  Evidence for a QW width crossover from double- to single-valley occupation has been shown for (001)AlAs wells \cite{vanKest,vandest}, as has evidence for single valley occupancy in wide (110)AlAs wells \cite{dasgupta2}.  Dynamic control of the valley degeneracy has been realized with uniaxially strained (001)AlAs QWs to induce valley degeneracy splitting \cite{shayeganreview, gun4, vaki1}.  Such studies can quantify interaction effects, calibrating valley strain susceptibility and valley effective mass \cite{gun2,gun4,gun3,shkolnikov}.  Quantum confinement to a one-dimensional multi-valley system has been achieved in cleaved-edge overgrown quantum wires \cite{moser1,moser2} and quantum point contacts \cite{gun1}. Novel interaction effects in QWs include anisotropic composite Fermion mass \cite{medini} and valley skyrmions, whereby electrons populate linear superpositions of two valleys at
once \cite{shkolskyrmions}.   Such interaction effects that result from exchange splitting of a perfect SU(2) symmetry\cite{rasolt} may prove useful in future quantum device applications, where the valley degree of freedom functions as a pseudospin.

Various formalisms have been developed to understand valley occupancy in a multi-valley system. AlAs QWs are slightly strained with respect to the GaAs substrate, so one must consider both quantum confinement and strain to estimate the electron subband energy and degeneracy of the different electron valleys, and certain growth symmetries also introduce in-plane or out-of-plane piezoelectric fields.
 Stern and Howard modeled two-dimensional (2D) confinement of electron valleys in Si for arbitrary crystal plane orientations neglecting strain effects \cite{stern}.
Van de Walle theoretically studied the absolute energy level for an unstrained semiconductor heterojunction, considering strain effects only for bulk systems \cite{vande}.
Smith {\em et al.}~calculated the strain tensor theoretically for the case of (001)- and (111)-oriented QW superlattices\cite{smithjap}.~Although Caridi and Stark derived the complete strain tensor for arbitrarily oriented substrates with cubic symmetry, they ignored a critical shear component which arises in the high Miller index directions under relaxation \cite{caridiapl}.
De Caro {\em et al.}~calculated the shear strain with commensurability constraint equations accurately only for the low index (100)-, (110)- and (111)-orientations \cite{decaro}.
Yang {\em et al.}~corrected the constraint equations and calculated piezoelectric fields arising from the shear strain components for the general case of a pseudomorphic film \cite{yangapl}.
Adachi calculated the strain tensor theoretically for the case of bulk and superlattice structures but did not address the case of single QW structures \cite{adachi}.
Hammerschmidt {\em et al.}~focus on the case of isotropic planar
strain arising in QWs but do not carry out a combined treatment of strain energy along with quantum confinement energy and how these affect valley degeneracies \cite{hammer}.
Theoretically, Rasolt proposed an understanding of the various continuous symmetries and symmetry-breaking in a multi-valley system in terms of an SU$(N)$ symmetry where $N$ is the valley degeneracy number\cite{rasolt}, but did not 
explicitly derive the vanishing intervalley exchange integral which underlies this theory. Material specific calculations of quantum confinement and strain in high-mobility Si and Ge structures are well studied for high-symmetry facets \cite{graf,boykPRB,boykAPL,nest,frie,virg,friePRB,gosw,nestPRB08,virgPRB09,fruc,schaffler} and recently, such calculations have been been made in other quantum confined systems as well \cite{kim,euar}.
However, a combined treatment of quantum confinement, strain, and piezoelectric fields  to determine valley degeneracy in QWs is lacking, especially for low symmetry facets.  
%

 To eventually model transport in such valley degenerate systems, one must also consider the extra scattering channel not present in single valley systems, namely intervalley scattering. It is therefore useful to introduce a $k$-space unit cell which permits visualization of momentum scattering events in a geometry that is natural to the quantum confinement direction. However, the standard depiction of a 2D Brillouin zone of a quantum confined valley-degenerate system \cite{stern}, projects all valleys to a single two-dimensional plane.   Such a depiction loses information about the out-of-plane momentum scattering component that was projected out, which is necessary to determine the full momentum scattering matrix element. Thus it is useful to develop a graphical representation for the unit cell in $k$-space that can clearly elucidate how valleys are coupled with both in-plane and out-of-plane momentum-scattering events. 
 %
%
%
%

The paper is organized as follows. In Section \ref{sec:exchange}, exchange energy calculations demonstrate why the valley index can function as a pseudospin index. In Section \ref{sec:analyt}, we develop the notation and formalism for determining valley degeneracy in multi-valley strained semiconductor QWs. The three subband energy components are defined -- kinetic energy, confinement energy and strain energy -- as are the two useful coordinate bases -- the conventional-cubic-cell basis 
and the transport basis -- 
as well as the coordinate transformation matrix which transforms between them.  Single-electron analytical solutions are provided to allow easy identification of characteristic energy scales for multi-valley systems. 
%
%
The explicit criteria for valley degeneracy in arbitrarily oriented biaxially strained QWs are defined in Section \ref{sec:vdegeneracy}.
Section \ref{sec:piezo} describes how shear strain can induce piezoelectric fields in the QW.
In Section \ref{sec:vsuc}, we introduce the valley-scattering unit cell -- a 3D primitive unit cell with the same volume as the Brillouin zone, but which shares the full symmetry of the reciprocal lattice vectors that lie in the Miller index plane.
Momentum-space illustrations provide intuition for visualizing intervalley scattering. 
Section \ref{sec:AlAs} applies the developed formalism to the case of AlAs QWs, whereby the 
projected in-plane transport masses, strain tensor, degeneracies, piezoelectric fields, and valley splitting energies for the specific cases of (001)-, (110)-, (111)- as well as the unconventional (411)-oriented QWs illustrate how valley degeneracies can be engineered. 
We review the crossover width 
calculation for double-to-single valley occupation in the (001)-, (110)-, and (411)-facet QWs. For the high-symmetry low Miller index orientations (001), (110) and (111) we calculate the biaxial Poisson ratio, and for the low-symmetry high Miller-index (411) facet there arises a shear component in the strain tensor and consequently a shear-to-biaxial strain ratio.
We calculate intervalley scattering ratios for valley-degenerate AlAs QWs on various facets in Section \ref{sec:vsucAlAs} and refer to the valley scattering unit cell for intuition.
We complete our analysis by comparing the standard 2D Bravais lattice valley representation to our valley cell representation. To generalize the formalism for calculating valley subband energies in arbitrarily oriented facets, we end in Section \ref{sec:Discussion} by adapting this notation for miscut samples \cite{eng}.

\section{Valley exchange integral and Valley index as a peudospin}\label{sec:exchange}
Just as electrons with different spins are distinguishable and have no exchange term in their interaction energy, electrons in different valleys can be shown to be {\em effectively} distinguishable from each other due to the vanishingly small intervalley exchange energy. The valley index can thus be treated as a pesudospin index. Rasolt\cite{rasolt} discusses symmetry-breaking in a multivalley system but does not explicitly derive this exchange integral. Previous work on Si quantum dots to study Kondo effect \cite{eto,hada,hadaprb} calculates the exchange integral only for the special case of identical spatial wavefunctions on Si quantum dots. For completeness, in this section we explicitly derive the inter- and intravalley exchange integrals for arbitrary wavefunctions to justify the valley index as a pseudospin, with exchange interaction proportional to $\delta_{\tau{\overline{\tau}}}$ where $\tau$ and $\overline{\tau}$ are valley indices.

An electron wavepacket within a single valley is described with a weighted integral over Bloch functions. The normalized wavepacket $\psi_{{\tau},\sigma}$ in valley $\tau$ is chosen to be in a spin polarized state $\sigma$ described by the spinor $\chi_\sigma$
\begin{equation}
\label{equ:blochwave}
\psi_{{\tau},\sigma}(\textbf{r}) = \frac{1}{\sqrt{V}}\int{d^3\textbf{k}~A_\textbf{k}~e^{i(\textbf{q}^\tau+\textbf{k})\cdot \textbf{r}}~u_{\textbf{q}^\tau+\textbf{k}}(\textbf{r})\chi_\sigma}~~~.
\end{equation}
$u_{\textbf{q}^\tau+\textbf{k}}(\textbf{r})$ is the component of the Bloch function periodic in the Bravais lattice, $A_\textbf{k}$ is the complex amplitude for a particular $\textbf{k}$, $\textbf{q}^\tau+\textbf{k}$ is the total crystal momentum where $\textbf{q}^\tau$ denotes  the center of the valley and $\textbf{k}$ denotes the additional small momentum deviation away from this valley center, and $V$ is the volume of the system.
Assuming $\textbf{k}$ is much smaller than the Umklapp vector for the lattice, and taking the envelope function approximation $u_{\textbf{q}^\tau + \textbf{k}} \cong u_{\textbf{q}^\tau}$, we obtain
\begin{eqnarray}
\label{equ:blochwave2}
\psi_{{\tau},\sigma}(\textbf{r}) &=& \frac{1}{\sqrt{V}}u_{\textbf{q}^\tau}(\textbf{r})e^{i\textbf{q}^\tau\cdot \textbf{r}}\chi_\sigma\int{d^3\textbf{k}~A_\textbf{k}e^{i\textbf{k}\cdot \textbf{r}}}\nonumber\\
&=& \frac{1}{\sqrt{V}}u_{\textbf{q}^\tau}(\textbf{r})e^{i\textbf{q}^\tau\cdot \textbf{r}}\chi_\sigma\phi(\textbf{r})
\end{eqnarray}
where $\phi(\textbf{r})$ is the slowly varying envelope function.
The Coulomb exchange energy integral between the two electron wavefunctions $\psi_{{\tau},\sigma}(\textbf{r})$ and $\overline{\psi}_{{\overline{\tau}},\overline{\sigma}}(\textbf{r})$ in valleys $\tau$ and $\overline{\tau}$ with coordinates $\textbf{r}_1$ and $\textbf{r}_2$ is given by
\begin{eqnarray}
\label{equ:exchange}
E_\mathrm{ex}^{ \tau, \overline{\tau}, \sigma, \overline{\sigma}}&=&\sum_{\sigma\overline{\sigma}}
\int\int{d\textbf{r}_1d\textbf{r}_2\frac{-e^2}{|\textbf{r}_1-\textbf{r}_2|}\psi_{\tau,\sigma}(\textbf{r}_1)\psi^*_{\tau,\sigma}(\textbf{r}_2)}\nonumber \\
& &\overline{\psi}_{{\overline{\tau}},{\overline{\sigma}}}(\textbf{r}_2){\overline{\psi}}^*_{{\overline{\tau}},{\overline{\sigma}}}(\textbf{r}_1)~~~.
\end{eqnarray}

Substituting Eq.~\eqref{equ:blochwave2} in Eq.~\eqref{equ:exchange}, we can show that two particles with different valley indices behave like distinguishable particles by virtue of their vanishing exchange integral,
\begin{eqnarray}
\label{equ:exchangedelta}E_\mathrm{ex}^{\tau, \overline{\tau}, \sigma, \overline{\sigma}}= \frac{\delta_{\sigma\overline{\sigma}}}{V^2}\int\int{d\textbf{r}_1d\textbf{r}_2~\phi(\textbf{r}_1)\phi^*(\textbf{r}_2){\overline{\phi}}(\textbf{r}_2){\overline{\phi}}^*(\textbf{r}_1)}\nonumber & &\\
 \frac{-e^2}{|\textbf{r}_1-\textbf{r}_2|}e^{i\textbf{Q}^{\tau\overline{\tau}}\cdot (\textbf{r}_2 - \textbf{r}_1)}u_{\textbf{q}^\tau}(\textbf{r}_1)u_{\textbf{q}^\tau}^*(\textbf{r}_2)\overline{u}_{\textbf{q}^{\overline{\tau}}}(\textbf{r}_2)\overline{u}_{\textbf{q}^{\overline{\tau} }}^*(\textbf{r}_1)& &\nonumber \\
\end{eqnarray}
with intervalley scattering wavevector
\begin{equation}
\textbf{Q}^{\tau\overline{\tau}} = \textbf{q}^{\tau} - \textbf{q}^{\overline{\tau}} ~~~,
\end{equation}
where the inner product of the spinors is given by ${\chi^\dag_{{\overline{\sigma}}}\chi_\sigma} = {\delta_{\sigma\overline{\sigma}}}$. 
This delta function denotes that the exchange integral vanishes when the two spin indices are different $\sigma \neq \overline{\sigma}$.
Analogously, if $\tau \neq \overline{\tau}$, $\textbf{Q}^{\tau\overline{\tau}}$ is of order an Umklapp vector, and the rapidly oscillating complex exponential makes the integral vanish for envelope functions larger than a few lattice constants. On the other hand if $\tau$ = $\overline{\tau}$, $\textbf{Q}^{\tau\overline{\tau}}$ = 0 and the exponential term is unity and the integral remains finite.
Therefore, the dependence of the exchange integral Eq.~\eqref{equ:exchangedelta} on valley index can be approximated with a second delta function to notate the vanishing exchange integral between different valleys
\begin{eqnarray}
\label{equ:equalvalleys}
E_\mathrm{ex}^{\tau, \overline{\tau}, \sigma, \overline{\sigma}}= \frac{\delta_{\sigma\overline{\sigma}}\delta_{\tau\overline{\tau}}}{V^2}\int{d\textbf{r}_1~\phi(\textbf{r}_1)\overline{\phi}^*(\textbf{r}_1)}|u_{\textbf{q}^\tau}(\textbf{r}_1)|^2 & & \nonumber \\
\int{d\textbf{r}_2~\phi^*(\textbf{r}_2)\overline{\phi}(\textbf{r}_2)}\frac{-e^2}{|\textbf{r}_1-\textbf{r}_2|}|u_{\textbf{q}^{{\tau}}}(\textbf{r}_2)|^2& .&
\end{eqnarray}
We note that Eq.~\eqref{equ:equalvalleys} cannot be simplified because the Coulomb potential can vary rapidly over small distances of order a lattice constant.

We conclude by virtue of this vanishing intervalley exchange interaction, that wavefunctions in different valleys can be treated like distinguishable particles, and the valley index is a valid pseudospin index. As was pointed out by Rasolt\cite{rasolt} the valley pseudospin constitutes an SU$(N)$ group where $N$ is the number of electron valleys.
%

\section{Valley subband energy}\label{sec:analyt}

Crystal symmetry dictates that multiple energy-degenerate valleys will occur whenever a local conduction band minimum exists away from the origin in momentum space (the $\Gamma$-point). When such a multi-valley system is then quantum confined in a layer with planar Miller indices $M = (h\,k\,l)$, the energy $E^{\tau}({\bf{k}})$ in the lowest subband in valley $\tau$ is
\begin{equation}
\label{equ:EnergySplitting}
 E^{\tau}({\bf{k}}) =  E_{0}^{\tau}({\bf k}) + T^{\tau}({\bf{k}}) + \Delta E^{\tau}
\end{equation}
where  ${\bf k}$ is the 2D in-plane momentum relative to the $\tau$-valley minimum, $E_{0}^{\tau}({\bf k})$ is the ground confinement energy, $T^{\tau}({\bf{k}})$ is the in-plane kinetic energy,  and $\Delta E^{\tau}$ is the strain induced energy shift caused by lattice mismatch of the QW with respect to the substrate. Note that the Miller index $M$ is not explicitly superscripted because it is common to all valleys.  To calculate these terms in Eq.~\eqref{equ:EnergySplitting}, we need to find the in-plane (parallel to the QW) and out-of-plane (confinement direction) components of the inverse mass tensor of the corresponding electron valley as well as the various strain tensor components in the QW.

We start by introducing two useful bases,~the conventional-cubic-cell (CCC) basis $\textbf{x} = (x,y,z)$ and the transport basis $\textbf{a} = (a,b,c)$, along with the
 coordinate transformation matrix ${\bf R}^M$ which transforms between them
\begin{equation}
\label{equ:rotationmatrix}
{\bf a = R}^M\textbf{x}~~~ .
\end{equation}
The CCC basis has the $x$, $y$, and $z$ axes aligned along the axes of the cubic cell of the reciprocal lattice of the crystal, Fig.\,\ref{fig:brill} (right). The transport basis is chosen with the $a$-$b$ plane parallel to the QW, and the $c$-direction perpendicular to this plane, Fig.\,\ref{fig:brill} (left).   Thus if the Miller index of the growth plane is ($h\,k\,l$), the perpendicular unit vector in the transport basis is $\bf{\widehat{c}}$ = ($h, k, l$)$\frac{1}{\sqrt{h^{2}+k^{2}+l^{2}}}$.  To uniquely define the $a$ and $b$ directions in the transport basis, we take $\bf{\widehat{a}}$ to be the in-plane unit vector with the lowest Miller index, and $\bf{\widehat{b}}$ the unit vector which maintains a right handed coordinate system $\bf{\widehat{a}} \times \bf{\widehat{b}}$ = $\bf{\widehat{c}}$. With this definition we obtain a unique set of axes.  These also define the components of the coordinate transformation matrix ${\bf R}^M$ whereby $\bf{\widehat{a}}$, $\bf{\widehat{b}}$, and $\bf{\widehat{c}}$ are the top, middle, and bottom rows of the coordinate transformation matrix ${\bf R}^M$.  In what follows, vectors and tensors which are expressed in the $\bf{x}$-basis are unprimed, and in the $\bf{a}$-basis they are primed.

The mass tensor is naturally expressed in the unprimed CCC \textbf{x}-basis, so by transforming it to the primed transport \textbf{a}-basis we can easily extract the in-plane and out-of-plane confinement masses for a given electron valley. The matrix inverse of the mass tensor for the $\tau$-valley in the $\textbf{x}$-basis is denoted by ${\bf w}^{\tau}$ = $({\bf m}^{\tau})^{-1}$. We assume parabolic bands so that the mass is independent of the wavevector {\bf{k}}.   Applying the transformation $({\bf {w}}'\,^{\tau})$ = ${\bf R}^M ({\bf w}^\tau)({\bf R}^M)^{-1}$ we obtain the inverse mass tensor of the $\tau$-valley in the primed transport ${\bf a}$-basis
\newcommand\ST{\rule[-1em]{0pt}{2.5em}}
\begin{equation}
\label{equ:MassTensor}
 ({\bf {w}}'\,^{\tau}) = \left[ \begin{array}{cc|c} {w}'\,^\tau_{aa}& {w}'\,^\tau_{ab}&~{w}'\,^\tau_{ac} \\ \ST {w}'\,^\tau_{ba}&{w}'\,^\tau_{bb} & ~{w}'\,^\tau_{bc}\\ \hline\ST {w}'\,^\tau_{ca}& {w}'\,^\tau_{cb}& ~{w}'\,^\tau_{cc} \end{array} \right].
\end{equation}
The upper left 2 $\times$ 2 sub-matrix ${\bf {w}}'\,^\tau_{2\times2}$ gives the inverse in-plane mass tensor and ${w}'\,^\tau_{cc}$ gives the inverse of the confinement mass perpendicular to the QW.

\subsection{Quantum confinement energy}\label{sec:Quantumconfinementenergy}
We can now define the first energetic term in Eq.~\eqref{equ:EnergySplitting}, the quantum confinement energy $E_{0}^{\tau}({\bf k})$,  using the coordinate transformation matrix and out-of-plane component of the inverse mass tensor in the primed $\bf{a}$-basis. This is calculated from the ground state energy solution of the 1D Schr\"odinger equation for a particle confined along ${c}$-direction with in-plane momentum \textbf{k},
\begin{equation}
\label{equ:SE}
\frac{-\hbar^2}{2} \frac{d}{dc}\left[ {w}'\,^\tau_{cc}(c)\frac{d\psi_{\bf k}(c)}{dc} \right] +V(c,{\bf k})\psi_{\bf k}(c) = E^{\tau}_{0}({\bf k})\psi_{\bf k}(c)
\end{equation}
where ${{w}'}_{cc}^\tau(c)$ is the diagonal component of the reciprocal mass tensor in the $\bf{a}$-basis. Due to the different reciprocal masses in the barrier (${\bf{w}}'\,_\textrm{B}^\tau$) and the well (${\bf{w}}'\,_\textrm{W}^\tau$) layers, wavefunction derivatives at the boundary $c_0$ must satisfy \cite{davies}
\begin{equation}
\label{equ:SE2}
{{w}'\,^\tau_{\textrm{B},cc}}\frac{d\psi_{\bf k}(c)}{dc}\biggr\rvert_{c = c_0^-} = {{w}'\,^\tau_{\textrm{W},cc}}\frac{d\psi_{\bf k}(c)}{dc}\biggr\rvert_{c = c_0^+}~~.
\end{equation}
%
The confinement potential $V(c,{\bf k})$ can be expressed as
\begin{equation}
\label{equ:barrier}
V(c,{\bf k}) = \left\{
\begin{array}{cll}
         0 & {\rm if} & |c| < W/2;\\
        V_0({\bf k}) & {\rm if} & |c| > W/2\end{array}
        \right.~~~,
\end{equation}
where the height of the potential barrier $V_0$ is given by
\begin{equation}
\label{equ:heightofbarrier}
V_0({\bf k}) = E_\mathrm{c,B}^\tau - E_\mathrm{c,W}^\tau + \frac{\hbar^2}{2} {\bf k} \cdot ({{\bf {w}}'\, }^\tau_\mathrm{2\times2, B} - {{\bf {w}}'\, }^\tau_\mathrm{2\times2, W})\cdot {\bf k}
\end{equation}
where $E_\mathrm{c,W}^\tau$ is the conduction band energy of the $\tau$-valley in the QW, $E_\mathrm{c,B}^\tau$ is the energy in the barrier, and $W$ is the well width. The last term accounts for differences in the {\em in-plane} effective mass as a ${\bf k}$-dependent barrier height.  This last term is often neglected in calculations under the assumption that ${\bf k}$ is small and the tensor mass difference ${{\bf {w}}'\,}^\tau_\mathrm{B} - {\bf {w}}'\,^\tau_\mathrm{W} $ is also small, making $E_{0}^{\tau}$ and $\psi(c)$ independent of in-plane momentum.

\subsection{Kinetic energy}\label{sec:Kineticenergy}
We can now calculate the second term in Eq.\,\eqref{equ:EnergySplitting}, the kinetic energy $T({\bf k})$, from the in-plane inverse mass tensor ${\bf {w}}'\,^\tau_{2\times2}$.  The in-plane inverse mass tensor is determined by solving the ground state from Eq.\,\eqref{equ:SE} and then determining a weighted average of the well and barrier inverse mass tensors given by
\begin{eqnarray}
\label{equ:weightedmass}
{\bf w}'\,^\tau_{2\times2} = 2 {\bf w'}_{{2\times2},\mathrm{W}}^\tau \int_0^{W/2}{|\psi(c)|^2}dc~+\nonumber \\
~2 {\bf w'}_{{2\times2},\mathrm{B}}^\tau \int_{W/2}^\infty{|\psi(c)|^2}dc~~~.
\end{eqnarray}
In most cases it is safe to assume the wide-well limit, whereby the effective mass is equal to that in the well material only. The kinetic energy is
 \begin{equation}
 \label{equ:KE}
 T^{\tau}({\bf{k}})= \frac{\hbar^2}{2} {\bf k}\cdot({{\bf {w}}'\,}^\tau_{2\times2})\cdot{\bf k}~~~.
 \end{equation}
 The projected in-plane effective masses can be solved from the determinant of this same sub-matrix
  \begin{equation}
 \label{equ:QE}
 \rm{det} ({{\bf {w}}'\,}^\tau_{2\times2} - \lambda {\bf I}) = 0~~~.
 \end{equation}
The eigenvalue solutions $\lambda_1$ and $\lambda_2$ directly give the reciprocal masses along the major and minor axes of the projected mass tensor.
The cyclotron mass is the same as the 2D density-of-states mass, and is given by
\begin{equation}
\label{equ:cycmass}
m_{\mathrm{2D}}^\tau=\mathrm{det}({\bf w'}\,^\tau_{2\times2})^{-1/2} = (\lambda_1\lambda_2)^{-1/2}~~~.
\end{equation}
The spin-degenerate two-dimensional energy density-of-states per unit area in the $\tau$-valley is $n_{\mathrm{2D}}^\tau = m_{\mathrm{2D}}^\tau/\pi \hbar^2$, so that the total energy density-of-states is given by a sum over all valleys
\begin{equation}
\label{equ:DOS}
 n_\mathrm{2D}(E) = \sum_\tau \frac{m_{\mathrm{2D}}^\tau}{\pi\hbar^2}~\Theta(E-E_0^\tau - \Delta E ^\tau)~~~.
\end{equation}
 where $\Theta(x)$ is the Heavyside step function.

\subsection{Strain energy}\label{sec:Strainenergy}
We can now define the final term of Eq.~\eqref{equ:EnergySplitting}, the strain energy $\Delta E^\tau$, in terms of the strain tensor and the coordinate transformation matrix. Following the notation of Van de Walle \cite{vande} and Herring and Vogt \cite{Herring}, the absolute energy shift $\Delta E^{\tau}$ of the $\tau$-valley for a homogeneous deformation described by the strain tensor $\boldsymbol\epsilon$ in the $\bf x$-basis is given by
\begin{equation}
\label{equ:delta}
 \Delta E^{\tau}= (\Xi_\mathrm{d}^\tau{\delta_{ij}} + \Xi_\mathrm{u}^\tau{\widehat{{q}}^{\tau}_{i}{\widehat{{q}}}^{\tau}_{j}} ) \epsilon^{ij}
\end{equation}
%
%
where  $\widehat{\bf{q}}^{\tau}$ is a unit vector in the direction of the $\tau$-valley, and $\Xi_\mathrm{d}^\tau$ and $\Xi_\mathrm{u}^\tau$ represent the deformation potentials due to a bulk dilation and a uniaxial deformation, respectively.
The average shift in the energy of the subband extrema is given by
\begin{equation}
\label{equ:delta1}
\Delta E_\mathrm{av}= (\Xi_\mathrm{d}^\tau + \frac{1}{3}\Xi_\mathrm{u}^\tau)\mathrm{\delta_{ij}}{\epsilon^{ij}}~=~a_\mathrm{c}\,\ \mathrm{\delta_{ij}}{\epsilon^{ij}}~~~~,
\end{equation}
where $a_\mathrm{c}$ is the hydrostatic deformation potential for the conduction band and $\mathrm{\delta_{ij}}{\epsilon^{ij}} = \mathrm{Tr}(\boldsymbol\epsilon)$ is the trace of the strain tensor $\epsilon_{xx}+\epsilon_{yy}+\epsilon_{zz}$. Often the relative strain energy shift $\Delta^{\tau}$ from the mean value is all that is needed to determine subband occupancy:
\begin{equation}
\label{equ:delta1}
\Delta^\tau= \Delta E^{\tau} - \Delta E_\mathrm{av} = ~\Xi_\mathrm{u}^\tau({\widehat{{q}}^{\tau}_{i}{\widehat{{q}}}^{\tau}_{j}}-\frac{1}{3}\mathrm{\delta_{ij}}){\epsilon^{ij}}~~~.
\end{equation}

To determine the strain tensor $\boldsymbol\epsilon$ 
it is easiest to explicitly determine its components in the ${\bf a}$-basis, which we will denote with ${\boldsymbol\epsilon}'$, and
 then apply a rotational transformation to express it in the ${\bf x}$-basis for use in Eqs.\,\eqref{equ:delta}-\eqref{equ:delta1}.
Heteroepitaxial QWs are biaxially strained relative to an unstrained substrate with a different lattice
  constant.  Thus the strain tensor ${{\boldsymbol \epsilon}'}$ is diagonal in the upper 2$\times$2 block
  ${{\epsilon}'}_{aa} = {{\epsilon}'}_{bb} = {{\epsilon}'}_{ \|}$, 
  where ${{\epsilon}'}_{ \|}$ is the in-plane strain component given by\cite{vande, kri}
 \begin{equation}
  \label{equ:inplanestrain}
 {\epsilon}'_ {\|}= \frac{a_{\mathrm {substrate}}-a_{\mathrm {layer}}}{a_{\mathrm {layer}}}~~~.
\end{equation}
Note that lattice matching to the unstrained substrate fixes $\epsilon'_{ab}=\epsilon'_{ba} = 0$.
The remaining components of the strain tensor are linearly proportional to the parallel strain ${\epsilon}'_ {\|}$ as follows:
\begin{equation}
\label{equ:straintensorcomplete}
 {\boldsymbol{\epsilon}}' = \left[ \begin{array}{ccc} 1& 0 &~D_1^M \\ \ST 0&1 & ~D_2^M\\ \ST D_1^M& D_2^M& ~-D_0^M \end{array} \right] {\epsilon}'_{\|}
\end{equation}
where $D_0^M = -\epsilon'_{cc}/\epsilon'_{aa} = -\epsilon'_{\|}/\epsilon'_{\bot}$ is the biaxial Poisson ratio (notated $D^M$ by Van de Walle \cite{vande}), and we define $D_1^M = \epsilon'_{ca}/\epsilon'_{aa}$ and $D_2^M = \epsilon'_{bc}/\epsilon'_{aa}$ as the shear-to-biaxial strain ratios.
%
%
\begin{table*}
 \begin{tabular}{|c||c|c|c|c|c||c|c|c||c|c||c|c||c|}
 \hline
\multirow{2}{*}  { }material  &   $D_0^{001}$ & $D_0^{110}$ & $D_0^{111}$ & $D_0^{411}$ &$D_2^{411}$ & $C_{11}$ & $C_{12}$ &$C_{44}$&$a_\mathrm{c}$ &$\Xi^\mathrm{X}_\mathrm{u}$&$m_\mathrm{l}$&$m_\mathrm{t}$&$e_{x,4}$\\
 &&&&&&\multicolumn{3}{|c||}{[GPa]}&[eV]&[eV]&$[m_{\mathrm e}]$&[$m_{\mathrm e}]$&[C/m$^2$]\\
\hline
\hline          AlAs& 0.854  &  0.616  & 0.550 & 0.775 & 0.176&125.0 & 53.4 & 54.2&2.54&6.11&1.1&0.20&-0.23\\
\hline          GaAs& 0.934  &  0.580  & 0.489 & 0.820& 0.250 & 122.1 & 56.6& 60.0&-0.16&8.61&1.3&0.23&-0.16  \\
\hline
\end{tabular}
\caption{Strain ratios $D_i^M$ for different facet orientations, elastic constants $C_{ij}$, deformation potentials $a_\mathrm{c}, \Xi^X_\mathrm{u}$, X-valley mass tensor components $m_\mathrm{l}, m_\mathrm{t}$, and piezoelectric coefficients $e_{x,4}$ for AlAs and GaAs \cite{stern,vurg,lan,vande,zunger}.}\label{Dvalues}
\end{table*}

The coefficients  $D_i^M$ can be derived for a crystal with cubic symmetry
 and arbitrary Miller index $M$ by minimizing the free energy of the layer in terms of its elastic constants \cite{vande,nyebook}. Using Hooke's law in the $\bf{x}$-basis, stress ($\boldsymbol\sigma$) and strain ($\boldsymbol\epsilon$) tensors are related by
\begin{equation}
\label{equ:strain1}
{{\sigma}}_{ij} =  {{c}_{ijkl}}~{{\epsilon}}_{kl}  ~~~~i,j,k,l \in \{x,y, z\}~~~.
\end{equation}
Here $\boldsymbol{c}$ is the fourth-rank elastic stiffness tensor in the $\bf{x}$-basis. For crystals with cubic symmetry
 $\boldsymbol{c}$  can be simplified using the Voigt notation \cite{voigt} into a matrix form
\begin{equation}
\label{equ:strain2}
 {\sigma}_{i}={C}_{ij}~{\epsilon}_{j}  ~~~~i,j = 1,2,...,6~~.
\end{equation}
The indices 1 through 6 denote $xx$, $yy$, $zz$, $yz$/$zy$, $zx$/$xz$ and $xy$/$yx$, respectively. For cubic materials most of the elements of the matrix ${C}_{ij}$ vanish and $C_{11} = C_{22} = C_{33}, ~C_{12} = C_{13} = C_{23}$ and $C_{44} = C_{55} = C_{66}$ simplifying the stress tensor $\boldsymbol\sigma$ to
\begin{equation}\label{strainmatrix}
    \left( {\begin{array}{c}
\sigma_1 \\
\sigma_2 \\
\sigma_3 \\
\sigma_4 \\
\sigma_5 \\
\sigma_6 \\
\end{array}} \right) =  \left( {\begin{array}{cccccc}
C_{11} & C_{12} & C_{12} & 0 & 0 & 0 \\
C_{12} &C_{11} &C_{12} &0 &0 &0 \\
C_{12} &C_{12} &C_{11} &0 &0 &0 \\
0 &0 &0 &C_{44} &0 &0 \\
0 &0 &0 &0 &C_{44} &0 \\
0 &0 &0 &0 &0 &C_{44}\\
\end{array}} \right) \left( {\begin{array}{c}
\epsilon_1 \\
\epsilon_2 \\
\epsilon_3 \\
\epsilon_4 \\
\epsilon_5 \\
\epsilon_6 \\
\end{array}} \right)~~~,
\end{equation}
where $\epsilon_1$ = $\epsilon_{xx}$, $\epsilon_2$ = $\epsilon_{yy}$, $\epsilon_3$ = $\epsilon_{zz}$, $\epsilon_4$ = $2\epsilon_{yz}$, $\epsilon_5$ = $2\epsilon_{zx}$ and $\epsilon_6$ = $2\epsilon_{xy}$. Table ~\ref{Dvalues} lists the values of $C_{11}, C_{12}$, and $C_{44}$ for AlAs and GaAs.
The strain tensor $\boldsymbol\epsilon$ and the elastic constants $c_{klmn}$ are used to obtain the free energy of isothermal elastic deformations of a medium\cite{hammer},
\begin{equation}
\label{equ:freeenergy}
F(\boldsymbol\epsilon)=\frac{1}{2}\sum_{klmn}c_{klmn}\epsilon_{kl}\epsilon_{mn}~~~.
\end{equation}
For a cubic crystal using the Voigt notation for the elastic constants, the elastic energy reduces to
\begin{eqnarray}
\label{equ:freeenergy2}
 F(\boldsymbol\epsilon)=\frac{C_{11}}{2}(\epsilon_{xx}^2 + \epsilon_{yy}^2 + \epsilon_{zz}^2)+ 2C_{44}(\epsilon_{xy}^2 + \epsilon_{xz}^2 + \epsilon_{yz}^2)\nonumber \\
 +~C_{12}(\epsilon_{xx}\epsilon_{yy} + \epsilon_{xx}\epsilon_{zz} + \epsilon_{yy}\epsilon_{zz})~~~.
\end{eqnarray}

The coefficients $D^M_i$ can be determined for a given crystal facet orientation by transforming the strain tensor of $\boldsymbol \epsilon'$ into the $\bf{x}$-basis with the rotation matrix
\begin{equation}
\label{equ:rotstrain}
{\boldsymbol \epsilon} = {({\bf R}^M)^{-1}({\boldsymbol {\epsilon}}' ){\bf R}^M}~~~,
 \end{equation}
and inserting these components into the elastic energy of Eq.~\eqref{equ:freeenergy2} and minimizing with respect to each $D^M_i$
\begin{equation}
\label{equ:freeenergy3}
F(\boldsymbol\epsilon) = F[{({\bf R}^M)^{-1}(\boldsymbol {\epsilon}' ){\bf R}^M}]
\end{equation}
\begin{equation}
\label{equ:minimumfreeenergy}
\frac{dF(\boldsymbol\epsilon)}{dD_i^M} = 0
\end{equation}
to deduce a set of three equations which can be simultaneously solved to give the $D^M_i$ values.
%
%
%
%
With the $D^M_i$ values solved for this material and this facet, the strain tensor in the $\bf x$-basis can now be used in Eqs.\,\eqref{equ:delta}-\eqref{equ:delta1} to deduce the strain energy contribution to each valley $\Delta E^\tau$.

As discussed in Section \ref{sec:AlAs}, for high-symmetry crystal orientations like (001), (110), and (111), there is no shear component in the strain tensor when expressed in the transport $\bf a$-basis  so that $D_1^M = D_2^M = 0$, whereas for (411) growth the $D_2^M$ coefficient is nonzero, and the direction of shear is defined by the vector  ${\boldsymbol { {\alpha}}}$ given by
\begin{equation}
\label{equ:defintionalpha}
{\boldsymbol { {\alpha}}} = D^M_1 {\bf {\widehat {a}}} + D^M_2 {\bf {\widehat {b}}}~~~.
\end{equation}

For AlAs and GaAs, we tabulate the nonzero $D_i^M$ in Table ~\ref{Dvalues} along with the elastic stiffness components in Voigt notation. For completeness, the Appendix provides a compact list of equations for determining the strain tensor for arbitrary substrate orientation. From the above analysis, we can now calculate all the terms in Eq.~\eqref{equ:EnergySplitting} for any valley and any QW orientation.

\section{Robust Valley Degeneracy and Crossover Valley Degeneracy}\label{sec:vdegeneracy}

Two valleys $\tau$ and $\overline{\tau}$ will be degenerate if
\begin{equation}
\label{equ:valleydegeneracy}
 {E}^{\tau} ({\bf k = 0}) = {E}^{\overline{\tau}}({\bf k = 0})~~.
\end{equation}
When this condition holds for $\tau$ and ${\overline{\tau}}$ regardless of the QW width, we call the valley degeneracy robust, and if it holds only for a specific well width $W_0$, we call it crossover degeneracy. The crossover degeneracy condition can arise only when the valley electrons with a higher confinement mass are strain-shifted to a higher energy than the valley electrons with a lower confinement mass.

Robust valley degeneracy can only occur if the strain energies and out-of-plane confinement masses are independently equal. The strain energy must satisfy
\begin{equation}
\label{equ:straindegeneracy}
 \Delta E^{\tau} = \Delta E^{\overline{\tau}}~~,
\end{equation}
for all well widths, which simplifies via Eq.~\eqref{equ:delta}  
to the $\bf{x}$-basis condition
\begin{equation}
\label{equ:valleydegeneracytensor}
 {\bf \widehat{q}}^{\,\tau}\cdot{\boldsymbol{\epsilon} }   \cdot  {\bf \widehat{q}}^{\,\tau}  =  {\bf \widehat{q}}^{\,\overline{\tau}}\cdot{\boldsymbol{\epsilon} }   \cdot  {\bf \widehat{q}}^{\,\overline{\tau}}
\end{equation}
or equivalently in the $\bf{a}$-basis,
\begin{equation}
\label{equ:valleydegeneracytensorabasis}
 {\bf \widehat{q'}}^{\,\tau}\cdot{\boldsymbol{\epsilon}' }   \cdot  {\bf \widehat{q'}}^{\,\tau}  =  {\bf \widehat{q'}}^{\,\overline{\tau}}\cdot{\boldsymbol{\epsilon}' }   \cdot  {\bf \widehat{q'}}^{\,\overline{\tau}}~~~.
\end{equation}
When $D^M_1 = D^M_2=0$, this condition for degeneracy simply reduces to
\begin{equation}
\label{equ:valleydegeneracyvector}
 |{\bf \widehat{q'}}^{\,\tau}\cdot{\bf \widehat{c}}| = |{\bf \widehat{q'}}^{\,\overline{\tau}}\cdot{\bf \widehat{c}}| ~~~,
\end{equation}
meaning that all valleys with the same polar angle from the $\bf c$-axis are degenerate.  For the more general case of arbitrary $D^M_i$, the full Eq.~\eqref{equ:valleydegeneracytensorabasis} has to be solved to determine valley degeneracy.  However, given the shear vector ${\boldsymbol {\mathrm {\alpha}}}$ in the plane of the QW as defined in Eq.~\eqref{equ:defintionalpha}, a special case can be defined, and two valleys $\tau$ and $\overline{\tau}$ are degenerate if they simultaneously satisfy Eq.~\eqref{equ:valleydegeneracyvector} and
\begin{equation}
\label{equ:valleydegeneracyvectoranydeformation}
|{\bf \widehat{q'}}^{\,\tau}\cdot{ {\widehat{\mathrm{\boldsymbol{\alpha}}}}}| = |{\bf \widehat{q'}}^{\,\overline{\tau}}\cdot{ {\widehat{\mathrm{\boldsymbol{\alpha}}}}}| ~~,
\end{equation}
or equivalently, if the valleys are mirror symmetric about the $c$-$\alpha$ plane.

The second condition for robust valley degeneracy is that the out-of-plane inverse masses are equal,
\begin{equation}
\label{equ:massdegeneracy}
 {w'}^{\tau}_{cc} = {w'}^{\, \overline{\tau}}_{cc}~~.
\end{equation}
guaranteeing equal confinement energies.  As long as the mass ellipsoid associated with each valley is oriented with its longitudinal mass parallel to the unit vector ${\bf \widehat{q}}^{\,\tau}$, as is usually the case, then the Eq.~\eqref{equ:massdegeneracy} is automatically satisfied under the same condition as Eq.~\eqref{equ:valleydegeneracyvector}.



\section{Piezoelectric effects on the quantum well potential}\label{sec:piezo}

In crystals which are inversion asymmetric like zinc blende crystals, strain may also produce piezoelectric fields, modifying the confinement potential $V(c,{\bf k})$, and resulting in shifts of the ground energies $E_0^\tau$ for the various valleys as solutions to the Schr\"oedinger equation, Eq.~\eqref{equ:SE}.  The out-of-plane piezoelectric field which results will not break robust valley degeneracy, but will shift the crossover degeneracy condition to a different well width, $W_0$.

In our treatment, we will calculate the piezoelectric field inside the QW and assume that any out-of-plane component will be canceled with an external gate bias, restoring the condition of the flat square well.  The piezoelectric field inside the QW and/or the barriers is proportional to the shear strain in the crystal basis \cite{adachi}

\begin{equation}
\label{equ:piezo}
\mathcal{E}_i = \frac{-1}{ \varepsilon_\mathrm{s}\varepsilon_0} e_{i,j}\,\epsilon_j ~~~.
\end{equation}

\noindent Here $\mathcal{E}_i$ lists the unprimed piezoelectric field components in the $\bf{x}$-basis indexed by $i = x,y,z$, $\varepsilon_0$ and $\varepsilon_s$ are the free-space and relative semiconductor dielectric constants, and $e_{i,j}$
represents the piezoelectric tensor with $j$ indices in Voigt notation. We assume that $\mathcal{E}_i$ is zero in the unstrained substrate. Recall that $\boldsymbol{\epsilon} = ({\bf R}^M)^{-1}(\boldsymbol{\epsilon'}) {\bf R}^M$ is the strain tensor derived from Eq.\,\eqref{equ:straintensorcomplete} but expressed in the unprimed $\bf x$-basis.~Even though the facet orientations (110) and (111) have no shear strain in the growth $\bf{a}$-basis, they do have shear strain in the crystal $\bf{x}$-basis. The (001) facet orientation, on the other hand does not have any shear strain. The strain in Eq.\,\eqref{equ:piezo} is represented in Voigt notation as for Eq.\,\eqref{strainmatrix} with $j = 1,2,...6$.  For zinc blende crystals, the only nonzero piezoelectric coefficients are $e_{x,4}=e_{y,5}=e_{z,6}$\cite{adachi} and we obtain
\begin{equation}\label{piezomatrix}
    \left( {\begin{array}{c}
\mathcal{E}_x \\
\mathcal{E}_y \\
\mathcal{E}_z \\
\end{array}} \right) =  \frac{-1}{\varepsilon_\mathrm{s}\varepsilon_0}\left( {\begin{array}{cccccc}
0 & 0 & 0 & e_{x,4} & 0 & 0 \\
0 & 0 & 0 & 0 & e_{x,4}  & 0 \\
0 & 0 & 0 & 0 & 0 & e_{x,4}  \\
\end{array}} \right) \left( {\begin{array}{c}
\epsilon_1 \\
\epsilon_2 \\
\epsilon_3 \\
\epsilon_4 \\
\epsilon_5 \\
\epsilon_6 \\
\end{array}} \right)~~~.
\end{equation}
Table \ref{Dvalues} lists the piezoelectric coefficients for GaAs and AlAs, and the coefficient for Al$_x$Ga$_{1-x}$As can be linearly interpolated according to the $x$-Al content\cite{adachi}. The general shear strain components in a QW are reproduced from Cadini and Stark \cite{caridiapl,GraysonAPLcomment} and simultaneously derived using Eq.\,\eqref{equ:rotstrain}, Eq.\,\eqref{equ:straintensorcomplete}, Eq.\,\eqref{equ:freeenergy3} and Eq.\,\eqref{equ:minimumfreeenergy}. It is useful to express the electric field in the primed growth $\bf a$-basis
\begin{equation}\label{piezoscalar}
    \left( {\begin{array}{c}
\mathcal{E'}_a \\
\mathcal{E'}_b \\
\mathcal{E'}_c \\
\end{array}} \right) =  \frac{-2 e_{x,4}}{\varepsilon_\mathrm{s}\varepsilon_0} ~{\bf R}^M
 \left( {\begin{array}{c}
\epsilon_{yz} \\
\epsilon_{zx} \\
\epsilon_{xy} \\
\end{array}} \right)~~~.
\end{equation}

 The $c$-component of the piezoelectric field $\mathcal{E}'_c$ will affect the electron subband energy by modifying the confinement potential $V(c,{\bf k})$ of Eq.\,\eqref{equ:barrier} both in the strained well and in the strained barrier
\begin{eqnarray}
\label{equ:barrierpiezo}
V(c,{\bf k}) = ~~~~~~~~~~~~~~~~~~~~~~~~~~~~~~~~~~~~~~~~~~~~~~~~~~~~~ \\
 \left\{
\begin{array}{cll}
         q\mathcal{E}'_\mathrm{c,W}\,c                                                                                 & {\rm if}~|c| < W/2;\\
        V_0({\bf k}) + q\mathcal{E}'_\mathrm{c,W}\,W/2 + q\mathcal{E}'_\mathrm{c,B}\,(c - W/2) & {\rm if}~c > W/2; \\
        V_0({\bf k}) - q\mathcal{E}'_\mathrm{c,W}\,W/2 + q\mathcal{E}'_\mathrm{c,B}\,(c + W/2) & {\rm if}~c < -W/2
        \nonumber, \end{array}
        \right.
\end{eqnarray}
where $\mathcal{E}'_{c,\mathrm{W}}$ and $\mathcal{E}'_{c,\mathrm{B}}$ are the out-of-plane $c$-components of the electric field in the well and barrier, respectively.  As pointed out by Adachi \cite{adachi} for (111) GaAs and AlAs, this piezoelectric field points from the (111)B surface (As-terminated) to the (111)A surface (Ga-terminated) when under positive $\bf c$-axis strain.

The piezoelectric field in the QW and barrier material alters both the quantum confinement energy and the wavefunction, and in the limit of weak piezoelectric fields the energy shift can be estimated to be $E_{\mathrm{pz}}^\tau =  \mathcal{E}'_cW/2$.  For example, in strained (111) AlAs on a GaAs substrate, the piezoelectric field is $\mathcal{E}'_c  = 8.794 \times 10^6\,\mathrm{V/m}$, so for a QW width of $W =$ 10 nm the piezoelectric energy shift would be around  $E_{\mathrm{pz}}^\tau =$ 43.97 meV.

For the present treatment, we reiterate our assumption that any nonzero out-of-plane component of piezoelectric field $\mathcal{E}'_c$ will be assumed to be canceled by an external gate voltage.

\begin{figure*}[h!t]
\includegraphics[width= 7in]{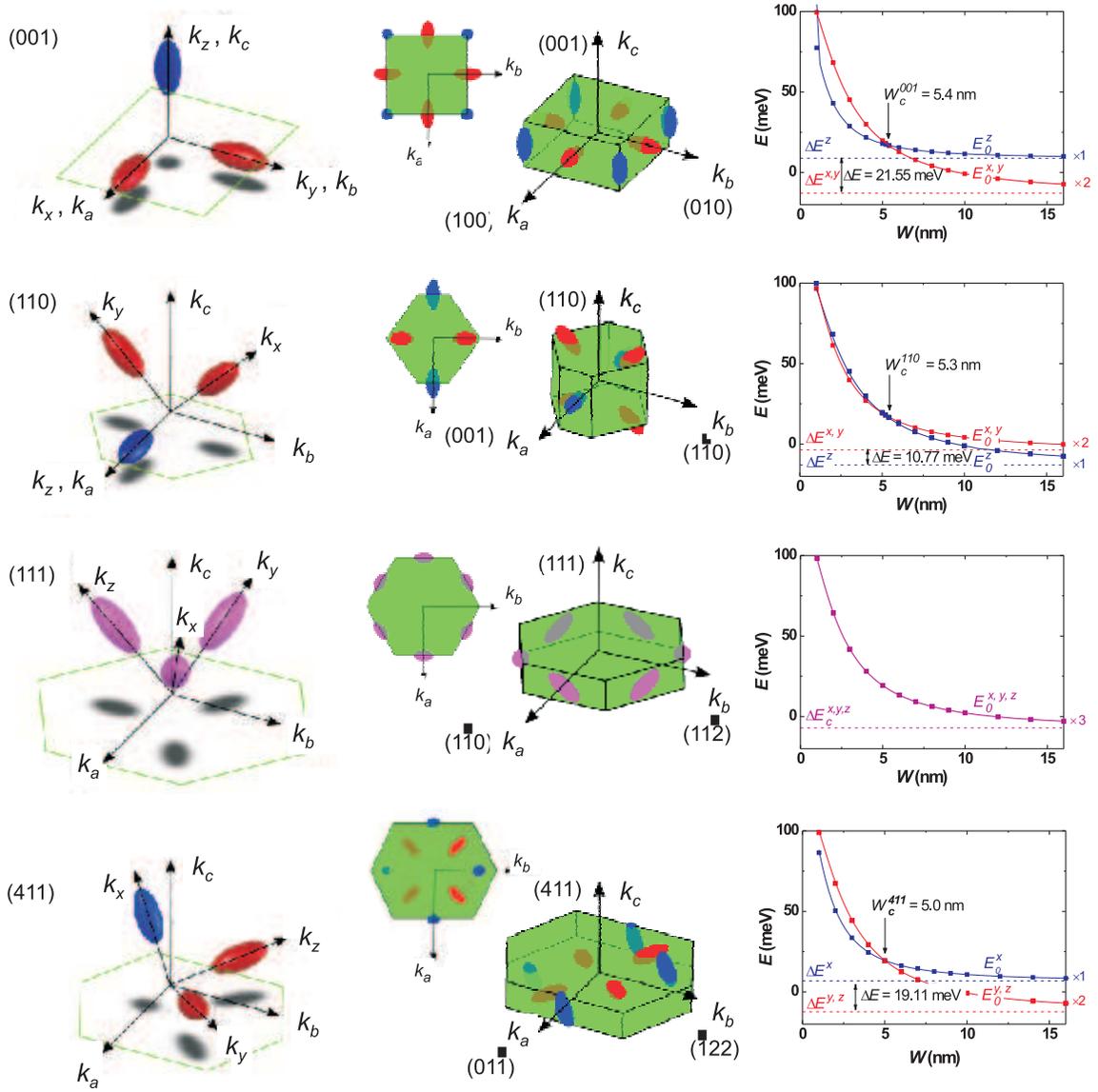}
	\caption{Depiction of the valley degeneracies for various QW orientations. Each row is labeled with its Miller index. Left column: 3D representation of the X-valleys showing both the CCC $k_x, k_y, k_z$ and the transport basis $k_a, k_b, k_c$. In all diagrams, the doubly-degenerate $X_{x,y}$ valleys are red, and the singly-degenerate $X_{z}$ valley is blue. For the (111)-oriented AlAs QW, the three degenerate $X_{x,y,z}$ valleys are purple. We note that the red and purple valleys exhibit robust valley degeneracy condition whilst the intersection point of the red and blue traces identifies the crossover degeneracy condition. The purple valleys satisfy the robust valley degeneracy condition as well as the crossover degeneracy condition trivially for all well widths. Center column: 2D projection and 3D representation of the valley scattering unit cell described in the text. Right column: Valley degenerate ground energy as a function of QW width for a QW barrier of Al$_{0.45}$Ga$_{0.55}$As at T = 4~K. Dashed lines represent the calculated strain energy shift of the respective electron valley relative to unstrained AlAs. The results of the heterostructure simulation software {\bf nextnano} are depicted with scatter plots, and are verified with the analytical calculations shown as continuous lines. The (111) and (411) calculations assume that the piezoelectric field has been canceled by a top gate to result in a flat QW.}
	\label{fig216aug}
\end{figure*}

\section{Valley Scattering Unit Cell}\label{sec:vsuc}
In addition to calculating valley subband energies and degeneracies, it is useful to visualize the valley orientations in three dimensions in order to map relevant intervalley scattering vectors. In this section, we lay down a general derivation of a 3D primitive unit cell which we call the valley scattering unit cell (VSC), that aids in identifying in-plane and out-of-plane intervalley scattering vectors. The proposed VSC is not a 3D Brillouin zone, but a simple unit cell defined to have the 2D symmetry of the Miller index plane of the QW, while preserving the total volume of the original 3D Brillouin zone. This cell is to be also distinguished from the standard depiction of a 2D Brillouin zone which neglects information about the out-of-plane momentum which is necessary for calculating momentum scattering matrix elements.

To define such a unit cell, we first identify the 2D subset of reciprocal Bravais lattice points which lie in the Miller index plane of the QW.  
The original 3D Bravais lattice is composed of parallel planes of this 2D sublattice which are displaced and separated by an interplanar wavevector $2\pi/d$. We now take the primitive unit cell of this 2D sublattice and extend it in the perpendicular direction by an amount $\pm \pi/d$, resulting in a primitive unit cell with the same volume as the original 3D Brillouin zone.  We call the resulting unit cell the valley scattering unit cell. We define coordinates for each
valley $\tau$ so that it lies within the VSC.

QW structures induce characteristically anisotropic scattering potentials, defined in the ${\bf a}$-basis as $V_\mathrm{s}(\emph{a},\emph{b},\emph{c})$, which can result in anisotropic intervalley scattering.  Scatterers  such as a single monolayer step in the sidewall of a QW or a miscut substrate with periodic sidewall steps are sharp on the order of a single lattice period and could in principle induce large momentum intervalley scattering in the $k$-space plane parallel to the QW.  

Recalling that the intervalley scattering vector is $\textbf{Q}^{\tau\overline{\tau}}$, we calculate the intervalley scattering matrix element,
\begin{eqnarray}
\label{scattering1}
v_{\tau\overline{\tau}} & = & \langle \psi_{\tau}(\textbf{r})|V_\mathrm{s}(\emph{a},\emph{b},\emph{c})|\overline{\psi}_{{\overline{\tau}}}(\textbf{r})\rangle \\
\label{scattering2}
& =& \langle \phi(\textbf{r})u_{\textbf{q}^\tau}(\textbf{r})|V_\mathrm{s}(\emph{a},\emph{b},\emph{c})e^{i\textbf{Q}^{\tau\overline{\tau}}\cdot (\textbf{a} + \textbf{b} + \textbf{c})}|\overline{\phi}(\textbf{r})u_{\textbf{q}^{\overline{\tau}}}(\textbf{r})\rangle \nonumber \\
\label{equ:scattering3}
& =& \int{d^3\textbf{r}\phi^*(\textbf{r})u^*_{\textbf{q}^\tau}(\textbf{r})V_\mathrm{s}(\emph{a},\emph{b},\emph{c})e^{i\textbf{Q}_\bot^{\tau\overline{\tau}}\cdot {\textbf{c}}}e^{i\textbf{Q}_\|^{\tau\overline{\tau}}\cdot{({\textbf{a}}+ {\textbf{b}})}}}\nonumber \\
& & \overline{\phi}(\textbf{r})u_{\textbf{q}^{\overline{\tau}}}(\textbf{r})
\end{eqnarray}

To emphasize the importance of the out-of-plane intervalley scattering vector, we write $\textbf{Q}^{\tau\overline{\tau}}$ in the exponential as the sum of an out-of-plane component $\textbf{Q}_\bot^{\tau\overline{\tau}}$ and the in-plane vector $\textbf{Q}_\|^{\tau\overline{\tau}}$.
If we use the standard planar 2D Brillouin zone view, we lose the representational importance of the out-of-plane scattering component.~Therefore, we choose to depict the VSC in all three dimensions. Specific examples will be shown in the following section which demonstrate how all possible intervalley scattering events can be easily identified in such a cell.


\section{Valley degeneracy and the valley scattering unit cell for AlAs QWs }\label{sec:AlAs}
 We apply the above analysis to the AlAs multi-valley system, determining the valley vectors, mass tensors, and strain tensors for the various growth directions, and then
  we calculate the valley subband energies as a  function of well width $W$ for each orientation.  The three degenerate conduction band electron valleys are composed of six
  half valleys located at the $X$-points of the Brillouin zone edge, with valley vectors ${\bf q}^{100} =  (2\pi/a) ~\widehat{\bf x}$, ${\bf q}^{010} =  (2\pi/a) ~\widehat{\bf y}$, and ${\bf q}^{001} =  (2\pi/a) ~\widehat{\bf z}$ for the $X_x, X_y,$ and $X_z$ points, respectively. The superscripts express $\tau$ as momentum-space directions after the notation of Van de Walle \cite{vande}. The AlAs electron valleys have anisotropic electron mass, with heavy longitudinal mass $m_{\mathrm l}$ = 1.1 $m_{\mathrm e}$ and light transverse mass $m_{\mathrm t}$ = 0.20 $m_{\mathrm e}$, respectively \cite {shayeganreview,gun2,gun3}. In the conventional crystal $\bf x$-basis, the mass tensor ${\bf m}^{\tau}$ is diagonal with ${m}^{\tau}_{ii} = m_{\mathrm l} $ for the mass component parallel to ${\bf q}^\tau$ and $m^\tau_{jj} = m_{\mathrm t} $ for the two transverse mass components repectively.  

Quantum confinement is created by sandwiching the AlAs QWs between aluminum gallium arsenide (Al$_x$Ga$_{1-x}$As) layers which have a high aluminium content $x >$ 0.4.
Properties of the barrier alloy are determined by interpolating between AlAs and GaAs, either linearly or with a bowing term where applicable.~We assume the reciprocal mass tensor in the ${\textrm{Al}}_{x}{\textrm{Ga}}_{1-x}{\textrm{As}}$ barrier layer to follow a linear interpolation for use in Eq.~\eqref{equ:SE} and \eqref{equ:SE2} \cite{footnoteadachi},
 \begin{equation}
\label{equ:SE1}
{{\bf{w}}'}^\tau_{{\textrm{Al}}_{x}{\textrm{Ga}}_{(1-x)}{\textrm{As}}} = (x){{\bf{w}}}'^\tau_{{\textrm{AlAs}}} + (1-x){{\bf{w}}}'^\tau_{{\textrm{GaAs}}}
\end{equation}
 For ${\textrm{Al}}_{x}{\textrm{Ga}}_{1-x}{\textrm{As}}$, $E_\mathrm{B}^\tau$ is given by the relation \cite{vurg}
 \begin{equation}
\label{equ:vegard}
E^\tau_{{\textrm{Al}}_{x}{\textrm{Ga}}_{1-x}{\textrm{As}}} = x E^\tau_{{\textrm{AlAs}}} + (1-x)E^\tau_{{\textrm{GaAs}}} - b x (1-x)~~~.
\end{equation}
 For ${\textrm{Al}}_{x}{\textrm{Ga}}_{1-x}{\textrm{As}}$ \cite{vurg}, the bowing term \cite{footnotebowing} $b^\tau$ is 0.055 eV for $E^\tau_\mathrm{GaAs}$ defined to be zero and $E^\tau_\mathrm{AlAs} = 0.259$ eV  where $\tau$ can be $X_x$, $X_y$ or $X_z$ respectively.

For all strain calculations, we assume AlAs to be~strained with respect to the GaAs substrate because in typical structures the thickness of the intervening buffer Al$_x$Ga$_{1-x}$As layer is below the critical strain relaxation thickness of 0.5 $\mu$m \cite{guti,weis}. Using Eq.~\eqref{equ:inplanestrain} we can calculate the in-plane strain from 
the lattice mismatch between the AlAs QW and the GaAs substrate
\begin{equation}
\label{equ:parallelstrainAlAs}
 {\epsilon}'_ {\|}= \frac{a_{\mathrm {GaAs}}-a_{\mathrm {AlAs}}}{a_{\mathrm {AlAs}}}~~~,
\end{equation}
where $a_{\mathrm {GaAs}}$ = 0.564177 {nm} (0.565325 {nm}) is the lattice constant of the GaAs substrate and $a_{\mathrm {AlAs}}$ = 0.565252 {nm} (0.566110 {nm}) is the lattice constant of the AlAs at 0 K (300 K)\cite{vurg}. The perpendicular strain component $\epsilon'_\bot$ is calculated using Eq.~\eqref{equ:straintensorcomplete} where $D^M_0$ is the biaxial Poisson ratio in the respective growth direction. 
The strain energy is determined from the absolute deformation potential for uniaxial strain at the $X$-point, $\Xi^X_\mathrm{u}$ = 6.11~eV in AlAs.

In the following subsections, we compare and contrast the calculated results for the (001)-, (110)-, (111)-, and (411)-oriented QWs, as summarized in Fig.~\ref{fig216aug}.  In the left panel of the figure, the transport basis vector $\bf{\widehat{c}}$ is taken along the growth direction. The differently colored ellipsoids distinguish the triple (purple), double (red), and single (blue) degenerate valleys, each of these satisfy the robust degeneracy condition given by Eqs.~\eqref{equ:valleydegeneracyvector} or \eqref{equ:valleydegeneracyvectoranydeformation}. For the (001), (110) and (411) orientations, the strain energy breaks the three fold valley degeneracy into twofold valleys and a single valley. The growth direction governs which of these valleys will have the lowest energy for a given range of well widths. For the (111) orientation, symmetry dictates threefold degenerate $X_{x,y,z}$ valleys for all well widths.

The right panel of Fig.~\ref{fig216aug} shows the results of the analytical simulation of $E_{0}^{\tau}$ which gives the valley subband energies as a function of the well width. The horizontal dashed lines indicate the pure strain splitting energy $\Delta E^{\tau}$, equivalent to the energy splitting in the asymptotic wide-well limit.   
The analytical results are confirmed by the semiconductor heterostructure simulation software {\bf nextnano} \cite{cknaak,nextnano}. 
For the crossover QW widths $W_0$ indicated for the (001), (110) and (411) orientations, all valleys satisfy the degeneracy condition of Eq.~\eqref{equ:valleydegeneracy} simultaneously. The intervalley scattering momentum vectors ${\bf{Q}}^{\tau\overline{\tau}}$ for all the facets are listed in Table ~\ref{alltable}.
\begin{table*}
\centering
\scriptsize{

\begin{tabular}{|c||c|c|c|c|}
\hline \hline
$M$ & \multicolumn{4}{|c|} {(001)~~~~~~~~~~~~~~~~~~~~~~~~~~~~ }\\\cline{1 -4}
         $\tau$& [100]  &  [010]  & [001]& \multicolumn{1}{|c|}{}  \\
\hline          ${\bf w'^\tau = w^\tau}$&$ \left[ \begin{array}{r@{}lr@{}l|r@{}l} 0&.91&0&&0&\\ 0&&5& &0&\\\hline  0&&0&&5& \end{array} \right] $   &   $\left[ \begin{array}{r@{}lr@{}l|r@{}l} 5&&0&&0&\\ 0&&0&.91&0&\\\hline  0&&0&&5&  \end{array} \right] $  & $ \left[ \begin{array}{r@{}lr@{}l|r@{}l} 5& & 0& & 0& \\ 0& & 5& & 0& \\ \hline 0& & 0& & 0&.91 \end{array} \right] $& \multicolumn{1}{|l|}{ $ \begin{array}{c}  {\bf{Q}}^{100,010}  \\ {\bf{Q}}^{010,001}  \\ {\bf{Q}}^{001,100}  \end{array}  $   $
\begin{array}{ccccc}  =( & ~1 & -1 & ~0 & )\frac{2\pi}{a}  \\ =( & ~0 & ~1 & ~0 & )\frac{2\pi}{a}  \\ =( & ~1 & ~0 & ~0 &)\frac{2\pi}{a}  \end{array}  $  } \\
 \hline      $m_{\mathrm{2D}}^\tau$ [$m_e$] & \multicolumn{2}{|c|}{0.469}   & 0.200 &  \multicolumn{1}{|c|}{} \\\cline{1 -4}
      $n_{\mathrm{2D}}^\tau$ [cm$^{-2}$meV$^{-1}$] & \multicolumn{2}{|c|}{3.92 $\times$ $10^{11}$}  & 8.36 $\times$ $10^{10}$& \multicolumn{1}{|c|}{}  \\

\hline \hline
$M$ & \multicolumn{4}{|c|} {(110)~~~~~~~~~~~~~~~~~~~~~~~~~~~~ } \\\cline{1 -4}
         $\tau$& [100]  &  [010]  & [001]& \multicolumn{1}{|c|}{}  \\
\hline          ${\bf w'^\tau}$&$ \left[ \begin{array}{r@{}lr@{}l|r@{}l} 5&&0&&0&\\ 0&&2&.95 &-2&.05\\\hline  0&&-2&.05&2&.95 \end{array} \right] $ & $ \left[ \begin{array}{r@{}lr@{}l|r@{}l} 5&&0&&0&\\ 0&&2&.95 &2&.05\\\hline  0&&2&.05&2&.95  \end{array} \right] $ & $ \left[ \begin{array}{r@{}lr@{}l|r@{}l} 0&.91 & 0& & 0& \\ 0& & 5& & 0& \\ \hline 0& & 0& & 5& \end{array} \right]$& \multicolumn{1}{|l|}{$  \begin{array}{c}  {\bf{Q}}^{100,010}  \\ {\bf{Q}}^{010,001}  \\ {\bf{Q}}^{001,100}  \end{array}  $   $  \begin{array}{r@{}lr@{}lr@{}l}  =( ~~0& & -1&.4142 & 0&~~~~~~~)\frac{2\pi}{a}  \\ =( ~~1& & 0&.7071 & -0&.7071  )\frac{2\pi}{a}  \\ =( -1& & 0&.7071 & 0&.7071 )\frac{2\pi}{a}  \end{array}  $ }  \\
\hline           $m_{\mathrm{2D}}^\tau$ [$m_e$] & \multicolumn{2}{|c|}{0.260}   & 0.469 &  \multicolumn{1}{|c|}{}  \\\cline{1 -4}
   $n_{\mathrm{2D}}^\tau$ [cm$^{-2}$meV$^{-1}$]& \multicolumn{2}{|c|}{2.18 $\times$ $10^{11}$}   & 1.96 $\times$ $10^{11}$& \multicolumn{1}{|c|}{} \\
\hline \hline
$M$ & \multicolumn{4}{|c|} {(111)~~~~~~~~~~~~~~~~~~~~~~~~~~~~ } \\\cline{1 -4}
          $\tau$& [100]  &  [010]  & [001]&  \multicolumn{1}{|c|}{}  \\
\hline          ${\bf w'^\tau}$& $\left[ \begin{array}{r@{}lr@{}l|r@{}l} 2&.95&-1&.18&-1&.67\\ -1&.18&4&.32 & -0&.96\\\hline  -1&.67&-0&.96&3&.64 \end{array} \right] $ & $ \left[ \begin{array}{r@{}lr@{}l|r@{}l}2&.95&1&.18&1&.67\\ 1&.18&4&.32 & -0&.96\\\hline  1&.67&-0&.96&3&.64 \end{array} \right] $ & $ \left[ \begin{array}{r@{}lr@{}l|r@{}l} 5& & 0& & 0& \\ 0& & 2&.27 & 1&.93 \\\hline 0& & 1&.93 & 3&.64 \end{array} \right]$  & \multicolumn{1}{|l|}{$  \begin{array}{c}  {\bf{Q}}^{100,010}  \\ {\bf{Q}}^{010,001}  \\ {\bf{Q}}^{001,100}  \end{array}  $   $  \begin{array}{r@{}lr@{}lr@{}l}  =( 1&.2247 & -2&.1213 & 0& )\frac{2\pi}{a}  \\ =( 2&.4494 & 0& & 0& )\frac{2\pi}{a}  \\ =( 1&.2247 &-0&.7071&0& )\frac{2\pi}{a}  \end{array}  $  } \\
\hline           $m_{\mathrm{2D}}^\tau$ [$m_e$] & \multicolumn{3}{|c|}{0.296}  &  \multicolumn{1}{|c|}{}  \\\cline{1 -4}
  $n_{\mathrm{2D}}^\tau$ [cm$^{-2}$meV$^{-1}$] & \multicolumn{3}{|c|}{3.72 $\times$ $10^{11}$}   &  \multicolumn{1}{|c|}{} \\
\hline \hline

$M$ & \multicolumn{4}{|c|} {(411)~~~~~~~~~~~~~~~~~~~~~~~~~~~~ } \\\cline{1 -4}
          $\tau$& [100]  &  [010]  & [001]&  \multicolumn{1}{|c|}{}  \\
\hline          ${\bf w'^\tau}$& $ \left[ \begin{array}{r@{}lr@{}l|r@{}l} 5&&0&&0&\\ 0&&4&.55 & 1&.29\\\hline  0&&1&.29&1&.36 \end{array} \right] $ & $ \left[ \begin{array}{r@{.}lr@{.}l|r@{.}l}2&95&-1&93&-0&68\\ -1&93&3&18 & -0&64\\\hline  -0&68&-0&64&4&77 \end{array} \right] $ & $ \left[ \begin{array}{r@{}lr@{}l|r@{}l} 2&.95 & 1&.93& 0&.68 \\ 1&.93 & 3&.18 & -0&.64 \\\hline 0&.68 & -0&.64 & 4&.77 \end{array} \right] $  & \multicolumn{1}{|l|}{$  \begin{array}{c}  {\bf{Q}}^{100,010}  \\ {\bf{Q}}^{010,001}  \\ {\bf{Q}}^{001,100}  \end{array}  $   $  \begin{array}{r@{}lr@{}lr@{}l}  =(  -0&.7071 & 0&.6667 & 0&.2357 )\frac{2\pi}{a}  \\ =( -1&.4141 & 0& & 0&~~~~~~~ )\frac{2\pi}{a}  \\ =( ~~0&.7071 &-0&.6667 & -0&.2357 )\frac{2\pi}{a}  \end{array}  $  } \\
\hline           $m_{\mathrm{2D}}^\tau$ [$m_e$] & 0.210& \multicolumn{2}{|c|}{0.420}   &  \multicolumn{1}{|c|}{} \\\cline{1 -4}
        $n_{\mathrm{2D}}^\tau$ [cm$^{-2}$meV$^{-1}$] & 8.77 $\times$ $10^{10}$   & \multicolumn{2}{|c|}{3.52 $\times$ $10^{11}$} &  \multicolumn{1}{|c|}{} \\
\hline

\end{tabular}}

\caption{Reciprocal mass tensors ${\bf w'}^\tau$ for the various valleys in the transport basis for (001)-, (110)-, (111)-, and (411)-oriented AlAs QWs.  Intervalley scattering vectors $Q^{\tau\overline{\tau}}$ are listed, as are the 2D density-of-states (cyclotron) masses relative to the free electron mass. The 2D density-of-states values $n_\mathrm{2D}^\tau$ are combined for robust-degenerate valleys.  }\label{alltable}
\end{table*}

\subsection{(001) AlAs}\label{subsec:001}

For the (001)-oriented QW, the crystal axes are identical to the growth axes, Fig.~\ref{fig216aug} (top row), thus the coordinate transformation matrix is the identity matrix ${\bf R}^{001} = {\bf I}$.
The inverse mass tensor follows trivially. The various mass parameters listed in Table ~\ref{alltable} are then used to calculate the respective energy terms in Eq.~\eqref{equ:EnergySplitting}.
The biaxial Poisson ratio derived from Section~\ref{sec:analyt} is given by\cite{vande} $D_0^M = 2\frac{c_{12}}{c_{11}}$, and the diagonal strain matrix has $D_1^{001}$ = $D_2^{001}$ = 0.
The strain tensor in the $\textbf{a}$-basis is
\begin{equation}
\label{equ:strain001}
 {\bf {\epsilon}}' = \left[ \begin{array}{ccc} 1& 0 &~0 \\ \ST 0&1 & 0\\ \ST 0& 0& ~-0.854 \end{array} \right]{\epsilon}'_{\|}~~~.
\end{equation}
The results of the Section~\ref{sec:analyt} analysis for (001)-oriented QWs including 2D density-of-states and cyclotron mass for each valley are provided in Table~\ref{alltable}.

The center panel of Fig.\,\ref{fig216aug} (001) depicts the VSC and its 2D projection. The 2D sublattice of coplanar reciprocal lattice points is square symmetric, with the $X_z$ valleys lying at the corners of the unit cell and the $X_{x,y}$ valleys at the edges. The resulting VSC illustrates that {\em all}~valleys can be connected by purely in-plane scattering events, which is not obvious from the usual depiction of (001) valleys. This can be verified in Table~\ref{alltable} where the $c$-component of the scattering vector ${\textbf{Q}}_\perp^{\tau \overline{\tau}}$ is zero for all intervalley scattering vectors.

As seen in Fig.~\ref{fig216aug} (001, right), strain alone as indicated by the horizontal dashed lines contributes an energy shift of $\Delta E^{100,010} = -11.62$\,meV for the doubly-degenerate $X_{x,y}$ valleys while raising the singly-degenerate strained $X_{z}$ valley by $\Delta E^{001}  = +9.93$\,meV, a net difference of $\Delta = 21.55$\,meV.  The ordering of the degeneracy changes at the crossover well width $W_{0}^\mathrm{001}$ = 5.4~nm.

Because (001) biaxial strain induces no shear component in the crystal basis, there are no piezoelectric fields for this facet.

\subsection{(110) AlAs}\label{subsec:110}
For the (110)-oriented QW in Fig.~\ref{fig216aug} (110, left), we apply the rules of axis identification described in Section ~\ref{sec:analyt}.  $\bf{\widehat{a}}$ lies along the lowest Miller index [001] direction, making the $X_z$ valley in-plane, and $\bf{\widehat{b}}$ along $[1\overline{1}0]$ completes the right handed coordinate system.  The coordinate transformation matrix ${\bf R}^{110}$ is
\begin{equation}
\label{R_{}110}
 {\bf R}^{110} = \left[ \begin{array}{ccc} 0 & 0 & 1\\\frac{1}{\sqrt{2}}&\frac{-1}{\sqrt{2}}&0 \\ \frac{1}{\sqrt{2}} & \frac{1}{\sqrt{2}} &0   \end{array} \right]~~~.
\end{equation}
The biaxial Poisson ratio derived from Section~\ref{sec:analyt} is given by\cite{vande} $D_0^M = \frac{c_{11}+3c_{12}-2c_{44}}{c_{11}+c_{12}+2c_{44}}$, and the diagonal strain matrix has $D_1^{110}$ = $D_2^{110}$ = 0.

The center panel of Fig.\,\ref{fig216aug} (110) depicts the VSC and its 2D projection. The strain tensor is diagonal in the $\textbf{a}$-basis and given by
\begin{equation}
\label{equ:strain110}
 {\bf {\epsilon}}' = \left[ \begin{array}{ccc} 1& 0 &~0 \\ \ST 0&1 & 0\\ \ST 0& 0& ~-0.616 \end{array} \right] {\epsilon}'_{\|}~~~.
\end{equation}
The results of Section~\ref{sec:analyt} analysis for (110)-oriented QWs including 2D density-of-states and cyclotron mass for each valley are provided in Table~\ref{alltable}.

The 2D sublattice which defines the VSC has a centered-rectangular symmetry. The $X_{z}$ valley lies in this plane and the $X_{x,y}$ valleys lie outside of the central plane.  In the VSC, it is clear that whereas the $X_{x,y}$ valleys can scatter amongst each other with purely in-plane scattering, the $X_z$ valley is isolated and requires an out-of-plane component for intervalley scattering. Table ~\ref{alltable} reflects this result since only the ${\bf{Q}}^{100,010}$ scattering vector has a zero $c$-component. 

As seen in Fig.~\ref{fig216aug}~(110, right), the singly-degenerate $X_{z}$ valley has the lowest strain energy $E^{001} = -12.94$\,meV whereas the doubly-degenerate strained $X_{x,y}$ valleys have an energy of $E^{100,010} =-3.55$\,meV due to the smaller compressive strain.  A smaller strain differential of $\Delta = 10.77$\,meV is observed between the valleys in this orientation as compared to the (001) orientation. We observe the valley degeneracy crossover at $W_0^{110}$ = 5.3~nm.

A purely in-plane piezoelectric field arises in (110) QWs due to the nonzero shear component of the strain tensor in the $\bf x$-basis.    When Eq.\,\eqref{equ:straintensorcomplete} is transformed to the $\bf x$-basis, the shear components take the form \cite{GraysonAPLcomment}
\begin{eqnarray}
\label{equ:straintensor110}
      & \epsilon_{xy} =  \frac{-C_{11}-2 C_{12}}{C_{11}+C_{12}+2 C_{44}} \epsilon'_{||}~~~; \\
      & \epsilon_{yz} = \epsilon_{zx} = 0 \nonumber
 \end{eqnarray}
leading to an electric field in the QW plane parallel to the [100] $a$-direction
\begin{equation}
\label{equ:efield110}
     \mathcal{E}'_a = \frac{-2 e_{x,4}}{\varepsilon_\mathrm{s} \varepsilon_0} \epsilon_{xy}~~~.
\end{equation}
This electric field $\mathcal{E}'_a = 7.942 \times 10^6~\mathrm{V/m}$ plays no role in quantum confinement so that the flat QW assumption remains valid, and the field is screened in-plane by the electrons in the QW within a Thomas-Fermi screening length of the sample edges at $\pm a$.


\subsection{(111) AlAs}\label{subsec:111}
For the (111)-oriented QW, $\bf{\widehat{a}}$ is chosen along the lowest Miller index $[1 \overline{1}  0]$  direction perpendicular to the growth axis $\bf{\widehat{c}}$, and none of the valleys lie within the plane.  $\bf{\widehat{b}}$ is chosen along the $[1 1 \overline{2}]$ direction to complete the right handed coordinate system \cite{rasolt}. The coordinate transformation matrix ${\bf R^{111}}$ is
\begin{equation}
\label{R}
 {\bf R^{111}} = \left[ \begin{array}{ccc} \frac{1}{\sqrt{2}}&\frac{-1}{\sqrt{2}}&0 \\\frac{1}{\sqrt{6}}&\frac{1}{\sqrt{6}}&\frac{-2}{\sqrt{6}}\\ \frac{1}{\sqrt{3}} & \frac{1}{\sqrt{3}} &\frac{1}{\sqrt{3}}   \end{array} \right]~~~.
\end{equation}
The biaxial Poisson ratio derived from Section~\ref{sec:analyt} is given by\cite{vande} $D_0^M = 2\frac{c_{11}+2c_{12}-2c_{44}}{c_{11}+2c_{12}+4c_{44}}$, and the strain matrix is diagonal with $D_1^{111}$ = $D_2^{111}$ = 0.
The strain tensor in the $\textbf{a}$-basis is
\begin{equation}
\label{equ:strain111}
 {\bf {\epsilon}}' = \left[ \begin{array}{ccc} 1& 0 &~0 \\ \ST 0&1 & 0\\ \ST 0& 0& ~-0.55 \end{array} \right] {\epsilon}'_{\|}~~~.
\end{equation}
The results of the Section~\ref{sec:analyt} analysis for (111)-oriented QWs including 2D density-of-states and cyclotron mass for each valley are provided in Table~\ref{alltable}. The (111) QW valleys remain threefold degenerate due to equal contributions of the strain and confinement energies for all electron valleys as seen in Fig.~\ref{fig216aug}~(111) left.

The center panel of Fig.\,\ref{fig216aug} (111) depicts the VSC and its 2D projection.  The planar sublattice is hexagonal, with six half-valleys located at the center of the hexagonal facets.  The resulting VSC shows that the valleys are all connected with coplanar scattering vectors, as seen in Table ~\ref{alltable} from the zero $c$-component of all ${\bf{Q}}^{\tau\overline{\tau}}$. Fig.~\ref{fig216aug}~(111, right), shows that all the valleys in this triply-degenerate system are equally strained to $\Delta E^{100,010,001} = -7.0$\,meV and the robust degeneracy condition given by Eq.~\eqref{equ:valleydegeneracy} is trivially satisfied for all well widths.

A purely out-of-plane piezoelectric field arises in (111) QWs due to the nonzero shear component of the strain tensor in the $\bf x$-basis.  When Eq.\,\eqref{equ:straintensorcomplete} is transformed to the $\bf x$-basis, the shear components take the form\cite{caridiapl,adachi}
\begin{equation}
\label{equ:straintensor111}
      \epsilon_{xy} = \epsilon_{yz} = \epsilon_{zx} =  \frac{-C_{11}-2 C_{12}}{C_{11}+2C_{12}+4 C_{44}} \epsilon'_{||}~~~,
 \end{equation}
leading to an electric field perpendicular to the QW plane parallel to the $c$-growth axis:
\begin{equation}
\label{equ:efield111}
     \mathcal{E}'_c = \frac{-2 e_{x,4}\sqrt{3}}{\varepsilon_\mathrm{s} \varepsilon_0} \epsilon_{xy}~~~.
\end{equation}
We will assume that this electric field $\mathcal{E}'_c = 8.794 \times 10^6~\mathrm{V/m}$ will be canceled by a gate voltage, so that the flat QW assumption remains valid.
\begin{figure}
\includegraphics[width= 3.5in]{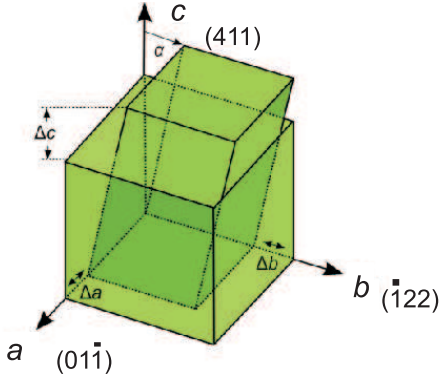}
	\caption{Illustration of shear strain for the (411)-oriented QWs due to the nonzero strain-to-shear ratio $D_1^{411}$.  $\Delta a$, $\Delta b$ and $\Delta c$ represent strain displacements along the respective vectors of the transport $\bf a$-basis, and $ \boldsymbol{\alpha}$ represents the shear vector in the plane of the QW, which for the (411)-oriented QWs is parallel to the $\bf b$-axis.
\label{411new}
}
\end{figure}
\subsection{(411) AlAs}\label{subsec:411}
To demonstrate the utility of the formalism described in this paper, we now extend our analysis to the less conventional (411) AlAs QW which has nonetheless shown important results in the literature \cite{depoorthesis,depoortereScience2000}. Applying the same rules of axis identification, $\bf{\widehat{a}}$ is chosen along the $[01\overline{1}]$ direction which is the lowest Miller index along the plane of the QW. $\bf{\widehat{b}}$ is chosen along the $[\overline{1}22]$  direction to complete the right-handed coordinate system. The coordinate tranformation matrix ${\bf R^{411}}$ is
\begin{equation}
\label{R411}
 {\bf R^{411}} = \left[ \begin{array}{ccc} 0&\frac{1}{\sqrt{2}}&\frac{-1}{\sqrt{2}} \\\frac{-1}{3}&\frac{2}{3}&\frac{2}{3}\\ \frac{4}{\sqrt{18}} & \frac{1}{\sqrt{18}} &\frac{1}{\sqrt{18}}\end{array} \right]~~~.
\end{equation}
Using the derivation for the strain ratios from Section~\ref{sec:analyt}, for the (411)-oriented QWs, we obtain $D_0^{411} = 0.775$ and $D_2^{411}  = 0.176$. The explicit relations for the biaxial Poisson ratios for AlAs (411)-QWs are
\begin{eqnarray}
\label{equ:D411}
D_0^{411}=  \nonumber \\
\frac{6(c_{11} + 2c_{12})(4c_{11} - 4c_{12} + 19c_{44})}{8c_{11}^2 - (16c_{12} - c_{44})(c_{12} + 2c_{44}) + c_{11}(8c_{12} + 145c_{44})} -1 \nonumber\\
D_2^{411}= \nonumber \\
\frac{-15 \sqrt{2} ({c_{11}}+2 {c_{12}}) ({c_{11}}-{c_{12}}-2 {c_{44}})} {8 {c_{11}}^2-(16 {c_{12}}-{c_{44}}) ({c_{12}}+2 {c_{44}})+{c_{11}} (8{c_{12}}+145 {c_{44}})} \nonumber~~~. \\
\end{eqnarray}

The results of Section~\ref{sec:analyt} analysis for (411) including 2D density-of-states and cyclotron mass for each valley are provided in Table\,\ref{alltable}. For this high-index facet, the strain tensor is nondiagonal in the $\textbf{a}$-basis and given by
\begin{equation}
\label{equ:strain411}
 {\bf {\epsilon}}' = \left[ \begin{array}{ccc} 1& 0 &~0 \\ \ST 0&1 & 0.176\\ \ST 0& 0.176& ~-0.775 \end{array} \right] {\epsilon}'_{\|}~~~.
\end{equation}
\begin{figure*}
\includegraphics[width= 7in]{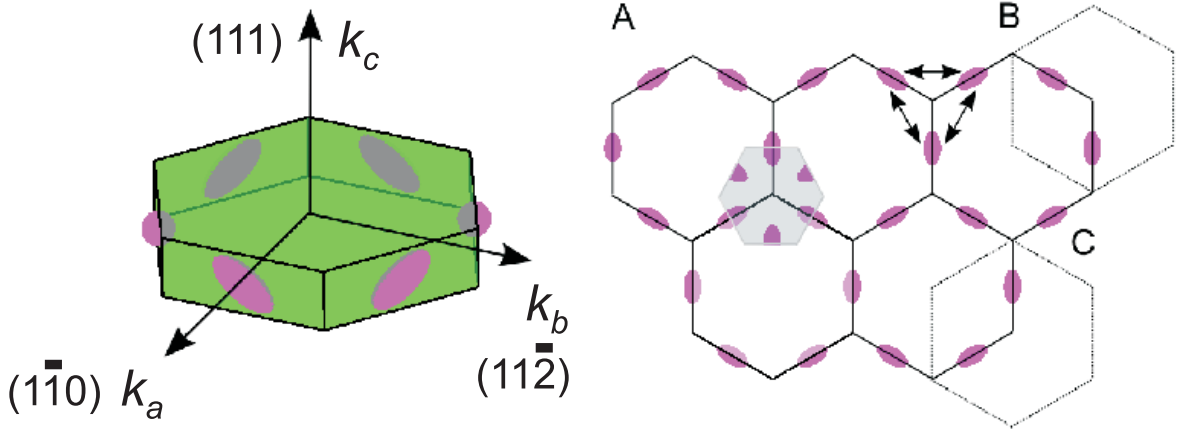}
	\caption{3D representation of the valley scattering cell for (111)AlAs as well as 2D lattice of such cells.
In AlAs, the valleys are located at the ${\bf{q}} =  (2\pi/a)(~\widehat{\bf x},~\widehat{\bf y},~\widehat{\bf z})$ The standard depiction of 2D Brillouin zone (small shaded hexagon)\cite{stern} projects together three identical but laterally and vertically displaced planar hexagonal lattices, translated according to the representative hexagonal cells A, B and C. The area of standard 2D Brillouin zone is one third that of the valley scattering cell since its height is three times that of the valley scattering cell.  Arrows illustrate the planar intervalley scattering vectors ${\bf{Q}}^{\tau\overline{\tau}}$ for (111)AlAs. 
\label{figSi}
}
\end{figure*}
The biaxial strain in the structure grown along [411] gives rise to nondiagonal components in the strain tensor. The effect of shear is depicted in Fig.~\ref{411new} by the shear vector ${\boldsymbol {\mathrm {\alpha}}}$ (Eq.~\eqref{equ:defintionalpha}) which lies in the plane of the QW and is parallel to the $b$-direction. The degeneracy condition takes into account the shear component $D_1^{411}$ and is given by Eq.~\eqref{equ:valleydegeneracyvectoranydeformation}.

Fig.~\ref{fig216aug}~(411, center) depicts the VSC and its 2D projection.
The 2D planar sublattice in Fig.\,\ref{fig216aug}~(411, center) has centered rectangular symmetry.

From Fig.~\ref{fig216aug} (411, right), we obtain a crossover width $W_{0}^{411}$ = 5.0~nm, and the valley occupation as a function of well width follows an analogous discussion as for the (110) case. However, we observe that the $X_{x}$ valley is singly-degenerate and the $X_{y,z}$ valleys are doubly-degenerate. The strain component of the $X_x$ valley energy is $\Delta E^{100} = 6.82$\,meV, and that of the $X_{y,z}$ valleys which satisfy the robust valley degeneracy condition is $\Delta E^{010} = -12.29$\,meV.  A strain differential of $\Delta = 19.11$\,meV is obtained. The two degenerate valleys have coplanar momentum scattering as seen in the vector ${\bf{Q}}^{010,001}$ which has zero $c$-component.

A piezoelectric field with both in-plane and out-of-plane components arises in (411) QWs due to the nonzero shear component of the strain tensor in the $\bf x$-basis.  When Eq.\,\eqref{equ:straintensorcomplete} is transformed to the $\bf x$-basis, the shear components take the form\cite{caridiapl}

\begin{equation}
\label{equ:straintensor411}
 \begin{array}{l}
\epsilon_{xy} = \epsilon_{zx} =\\
 \frac{ -(C_{11}+ 2 C_{12})(17 C_{11} - 17 C_{12} + 2 C_{44})}{8 C_{11}^2 - (16 C_{12} - C_{44}) (C_{12} + 2 C_{44}) +
   C_{11} (8 C_{12} + 145 C_{44})} \epsilon_{||} \\
 \epsilon_{yz} =\\
 \frac{-(C_{11} + 2 C_{12}) (8 C_{11} - 8 C_{12} - 7 C_{44})}{8 C_{11}^2 - (16 C_{12} - C_{44}) (C_{12} + 2 C_{44}) +
   C_{11} (8 C_{12} + 145 C_{44})} \epsilon_{||} ~~~,\\
 \end{array}
 \end{equation}
leading to an electric field with components both in the QW $a$-$b$ plane as well as parallel to the $c$-growth axis:
\begin{equation}
\label{equ:efield411}
     \boldsymbol{\mathcal{E}}' = - \frac{2 e_{x,4}}{\varepsilon_\mathrm{s} \varepsilon_0}
     \left( \begin{array}{c}
      -\frac{1}{\sqrt{2}} \epsilon_{xy}  + \frac{1}{\sqrt{2}}  \epsilon_{zx} \\
       +\frac{4}{3} \epsilon_{xy}  - \frac{1}{3} \epsilon_{yz} \\
       \frac{2}{\sqrt{18}} \epsilon_{xy}  + \frac{4}{\sqrt{18}} \epsilon_{yz} \\
     \end{array} \right)~~~.
\end{equation}
We will assume that the out-of-plane piezoelectric field component $\mathcal{E}'_c = 1.783 \times 10^6 \mathrm{V/m}$ will be cancelled by a gate voltage so that the square QW assumption remains valid.
The piezoelectric field component $\mathcal{E}'_a$ is zero, and the in-plane component ${\mathcal{E}}'_b = 3.761 \times 10^6$~V/m will be screened at the sample edges by the electrons in the QW.
%

\section{The valley scattering unit cell and Anisotropic intervalley scattering for AlAs QWs}\label{sec:vsucAlAs}

This section first discusses the differences between the VSC and standard depiction of the 2D Brillouin zone, and then determines the inter- to intravalley scattering ratio for valleys near crossover degeneracy.
In the VSC description for the case of (111)AlAs, Fig.\,\ref{figSi} shows the (111)AlAs VSC, as well as the 2D Bravais lattice of these cells.  For comparison with this hexagonal unit cell, the standard 2D Brillouin zone is plotted as a smaller grey hexagon within this lattice \cite{stern}. The hexagons A, B and C are intended to illustrate the relative positions of three identical planar hexagonal lattices, displaced from each other. When these three layers are stacked in sequence ABCABC¡­ out of the plane we recover the full 3D reciprocal lattice.  Whereas the standard 2D Brillouin zone is defined from the overlay of all three layers, we define the VSC from the 2D Brillouin zone of a single layer only. The area of the hexagonal VSC for (111) is thus three times larger than that of the traditional hexagonal unit cell. The coplanar intervalley momentum scattering vectors are shown with arrows in the VSC.

The VSC is helpful in depicting intervalley scattering events, so we will discuss the two main sources of single-particle scattering for AlAs QWs, namely  interface roughness and alloy disorder scattering in the barrier. We find below that near valley degeneracy, interface roughness scattering does not result in any significant intervalley scattering, and in the presence of alloy scattering in the barrier walls, the scattering rate for crossover intervalley scattering is significantly repressed relative to robust intervalley scattering, and this suppression factor is calculated.

We assume low enough temperatures such that acoustic and optical phonons can be neglected. Elastic intervalley scattering can only occur if the valley splitting energy is less than the Fermi energy at low temperatures, $| E^\tau-E^{\overline{\tau}} | < E_\mathrm{F}$, or the thermal energy at high temperatures $| E^\tau-E^{\overline{\tau}} |  < k_\mathrm{B}T$. Two independent processes can in principle result in elastic intervalley scattering: interface roughness scattering and alloy disorder scattering due to wavefunction penetration into the barrier alloy.

Interface roughness leads to local changes in the well width $\Delta(\textbf{r})$ at an in-plane position $\textbf{r} = (a,b)$, and the resulting fluctuation in the quantization energy causes a scattering potential\cite{sakaki}.~The roughness is characterized by the rms average displacement of the interface $\Delta$ and the roughness correlation length along the interface plane $\Lambda$, and is expressed as the Gaussian autocorrelation function \cite{ando}
 \begin{equation}
\label{equ:autocorrelation}
<\Delta(r)\Delta(r')> = \Delta^2\textrm{exp}\left(-\frac{{\mid r -r'\mid}^2}{\Lambda^2}\right)~~~,
\end{equation}
where $<...>$ means an ensemble average. The Fourier transform of the roughness autocorrelation at the intervalley scattering vector $\textbf{Q}^{\tau\overline{\tau}}$ is
 \begin{equation}
\label{equ:autocorrelation2}
<\mid\Delta_{\textbf{Q}^{\tau\overline{\tau}}}\mid^2> = \pi(\Delta^2{\Lambda}^2)\textrm{exp}\left[-\frac{({\Lambda\textbf{Q}^{\tau\overline{\tau}})^2}}{4}\right]~~~.
\end{equation}
Following Quang {\em et al.} \cite{quang1, quang2} the scattering potential $U_\mathrm{RS}$ for the QW with a square potential and a well width $W$ is given by

\begin{equation}
\label{equ:scatteringmatrix element}
<\mid U_\mathrm{RS}^{\tau\overline{\tau}}\mid^2>  =  A\left(\frac{\hbar^2{w}'\,^\tau_{\textrm{W},cc}}{2W^3}\right)~ <\mid\Delta_{\textbf{Q}^{\tau\overline{\tau}}}\mid^2>~~~,
\end{equation}
where ${w}'\,^\tau_{\textrm{W},cc}$ is the inverse confinement mass in the well perpendicular to the interface and $A$ is a unitless normalization constant. Under optimal epitaxial conditions of slip-step growth, the AlAs roughness disorder potential will have correlation lengths of tens of lattice periods  $\Lambda = Na$.  Since the intervalley scattering vector $\textbf{Q}^{\tau\overline{\tau}}$ is of order a Brillouin zone boundary $\pi/a$ and the intravalley scattering vector is of order zero, the ratio of intervalley scattering to intravalley scattering is of order $\textrm{exp}(-N^2)$, which even for a conservative estimate of $N=5$ lattice periods of autocorrelation already yields a suppression of more than 10 orders of magnitude for intervalley scattering.  Thus intervalley surface-roughness scattering is significantly suppressed with respect to intravalley scattering.

Alloy scattering is the other possible mechanism for elastic intervalley scattering.  The standard model for alloy disorder scattering \cite{quang1, murphy} with variable alloy composition $x(z)$ is described by the autocorrelation function
\begin{equation}
\label{equ:scatteringmatrix elementalloyscattering}
<\mid U_\mathrm{AD}^{\tau\overline{\tau}}\mid^2> = \frac{a_0^3 V_0^2}{8}\int_\frac{W}{2}^\infty{\psi_{\tau}^2(c)x(c)[1-x(c)]\psi_{\overline{\tau}}^2(c)dc}~~~,
\end{equation}
where $a_0$ is the in-plane lattice constant, and $V_0$ is the spatial average of the fluctuating alloy potential over the alloy unit cell. In AlAs QWs, only tails of the wavefunction have an alloy content in the Al$_x$Ga$_{1-x}$As barrier. One notes that the wave functions of electrons in valleys with a light confinement mass penetrate to a greater extent in the barrier and will have a significantly larger alloy scattering potential as compared to the heavy confinement mass. Assuming valley degeneracy for all valleys and plane wave solution for the electron wavefunction
 in the QW barrier, the scattering matrix element can be written as
\begin{eqnarray}
\label{equ:scatteringmatrix elementalloyscattering2}
<\mid U_\mathrm{AD}^{\tau\overline{\tau}}\mid^2> = \,\,\,\,\,\,\,\,\,\,\,\,\,\,\,\,\,\,\,\,\,\,\,\,\,\,\,\,\,\,\,\,\,\,\,\,\,\,\,\,\,\,\,\,\,\,\,\,\,\,\,\,\,\,\,\,\,\,\,\,\,\,\,\,\,\,\,\,\,\,\,\,\,\,\,\,\,\,\,\,\,\,\,\,\,\,\,\,\,\,\,\,\ \nonumber\\
\frac{a_0^3 V_0^2}{8}\int_\frac{W}{2}^\infty{{A_{\tau}^2 e^{-2\kappa^{\tau}c}}x(c)[1-x(c)]A_{\overline{\tau}}^2 e^{-2\kappa^{\overline{\tau}}c}dc}\nonumber\\~~~,
\end{eqnarray}
where $A_{\tau} = |\psi_\tau(W/2)|$ and $A_{\overline{\tau}} = |\psi_{\overline{\tau}}(W/2)|$ are the boundary values of the wavefunction at the barrier, obtained by solving for boundary conditions of the square wave potential.

We get a quantitative estimate of the ratio of the alloy scattering $\gamma_\times$ between crossover-degenerate valleys labelled $\tau_1$ and $\tau_3$ compared to the scattering $\gamma_\|$ between robust-degenerate valleys labeled $\tau_1$ and $\tau_2$ in the Al$_x$Ga$_{1-x}$As barrier by calculating
\begin{eqnarray}
\label{equ:scatteringmatrix elementalloyscattering2}
\frac{<\mid U_\mathrm{AD}^{\tau_1\tau_3}\mid^2>}{<\mid U_\mathrm{AD}^{\tau_1 \tau_2}\mid^2>} = \frac{\gamma_\times}{\gamma_\|} = \,\,\,\,\,\,\,\,\,\,\,\,\,\,\,\,\,\,\,\,\,\,\,\,\,\,\,\,\,\,\,\,\,\,\,\,\,\,\,\,\,\,\,\,\,\,\,\,\,\,\,\,\,\,\,\,\,\,\,\,\,\,\,\,\,\,\,\,\,\,\,\,\,\,\,\,\,\,\,\,\,\,\,\ \\
\frac{\int_\frac{W}{2}^\infty{{A_{\tau_1}^2 e^{-2\kappa^{\tau_1}c}}\,x(c)[1-x(c)]A_{\tau_3}^2 e^{-2{\kappa}^{\tau_3}c}dc}}{\int_\frac{W}{2}^\infty{{A_{\tau_1}^2 e^{-2{\kappa}^{\tau_1}c}}\,x(c)[1-x(c)]A_{\tau_2}^2 e^{-2{\kappa}^{\tau_2}c}dc}}\nonumber
\end{eqnarray}

\noindent where $c$ is the direction of integration out-of-plane.  For the case of (001)-, (110)-, and (411)-oriented AlAs QWs, $\frac{\gamma_\times}{\gamma_\|}$ = 0.0012, 0.0691 and 0.0235 respectively. The valley with the larger confinement mass penetrates less into the barrier region and results in a smaller scattering integral ratio.
The VSC makes it easy to visually identify this anisotropic scattering possibility along the various orientations.

Electron-electron intervalley scattering is outside the scope of this article and will be treated elsewhere.  It is expected to be small since the Fermi wavevector $k_\mathrm{F}$ is much smaller than the intervalley scattering vector \cite{wadedegottardi} $Q^{\tau\overline{\tau}}$.

\section{Valley degeneracy in miscut samples}\label{sec:Discussion}
In this final section, we show how the analysis of Section \ref{sec:analyt} can be applied to miscut samples to determine the projected masses, ground state energies, and strain energies.~Miscut samples are prevalent in the literature, for example (111) GaAs/AlAs growth has been shown to have superior morphology with a miscut angle from 0.5$^\circ$ to 4$^\circ$ and in general, intentional miscuts improve growth quality due to slip-step growth \cite{Vina, hayakawa, chin, tsutsui}. In Si, recently investigated hydrogen terminated (111)Si miscut surfaces have shown very high mobility \cite{engapl} and the wafer miscut is expected to break the valley degeneracy \cite{eng, kharche}.

Miscut samples are characterized by two angles. 
One angle $\phi$ designates the azimuthal angle in the $a$-$b$ plane relative to the $\bf a$-axis towards which the plane is tilted, often expressed as an in-plane Miller index tilt direction $T$.  The other angle $\theta$ designates the polar tilt angle relative to the surface normal $\widehat{\bf c}$.
Experimentally, the angles can be deduced from atomic force microscope images of surface monolayer steps, where $\phi$ is oriented perpendicular to the steps, and $\theta = \mathrm{atan}(a/2L)$ is deduced from knowing the monolayer thickness $a/2$ and the average width of the monolayer steps $L$.  We define a new coordinate transformation matrix ${{\bf R}^{M*}}$ for the miscut samples, by introducing the azimuthal and polar rotations ${\textbf{R}}_\phi$ and ${\textbf{R}}_\theta$, respectively:

\begin{equation}
\label{Rvarphi}
 {{\bf {R}}_{T}}={\bf R_{\phi}} = 
 \left[ \begin{array}{ccc} \mathrm{cos}\phi& -\mathrm{sin}\phi & 0\\ \mathrm{sin}\phi & \mathrm{cos} \phi & 0 \\ 0 & 0 &1  \end{array} \right]
\end{equation}
%
\begin{equation}
\label{Rphi}
  {\bf R_{\theta}} = 
  \left[ \begin{array}{ccc} \mathrm{cos}\theta&0& \mathrm{sin}\theta \\0&1&0\\ -\mathrm{sin}\theta&0 &\mathrm{cos}\theta    \end{array} \right]~~~.
\end{equation}

We obtain a new coordinate transformation matrix ${{\bf R}^{M}_{{\theta},{\phi}}}$ for calculating the transport parameters and the strain tensor analysis of miscut samples
\begin{equation}
\label{Rmiscut}
 {{\bf R}^{M}_{{\theta},{\phi}}} =    ({\bf R}_\phi {\bf R}_\theta {\bf R}^{-1}_\phi) {\bf R}^M~~~.
\end{equation}
%
Note that a VSC cannot be defined for miscut samples, but only for those with a pure Miller index.

We outline here the procedure for determining valley subband energies under miscut.  First, the transformation matrix ${{\bf R}^{M}_{{\theta},{\phi}}} $ from Eq.~\eqref{Rmiscut} determines the inverse mass tensor of each valley with Eq.~\eqref{equ:MassTensor}, and the out-of-plane inverse mass tensor component can be used in Eq.~\eqref{equ:SE} to determine the ground confinement energies of each valley. The in-plane inverse mass tensor is then calculated with Eq.~\eqref{equ:weightedmass} and used in Eq.~\eqref{equ:KE} to obtain the kinetic energy. The stress tensor can be determined by applying ${{\bf R}^{M}_{{\theta},{\phi}}} $ in Eq.~\eqref{equ:rotstrain}, and the strain ratios $D_i^M$ can be determined by minimizing the elastic energy with Eqs.~\eqref{equ:freeenergy2}-\eqref{equ:freeenergy3}.  Finally, the strain energy shift can be deduced from Eq.~\eqref{equ:delta} and added to the ground quantum confinement energies and kinetic energies to arrive at the final valley energies.
Any piezoelectric fields can be determined by transforming the strain tensor to the unprimed frame $\boldsymbol{\epsilon}$ and inserting the appropriate shear components into Eq.~\eqref{piezoscalar}.



\section{Conclusion}\label{sec:Conclusion}
In conclusion, we review the advantages and limitations of the formalism we have developed for valley subband energy calculations. We first derived how the valley index is a valid pseudospin index in a multi-valley system. A key element of our model is the definition of two relevant bases: the conventional crystal $\bf x$- and the transport $\bf a$-basis. We use this notation to find the projected in-plane and out-of-plane effective masses, the electron subband energy and degeneracy, piezoelectric fields, and the scattering vectors.

 There are five competing energy scales in multi-valley QWs namely in-plane kinetic energy, confinement energy, strain energy, piezoelectric energy, and Fermi energy, all of which influence valley occupancy. Strain in the QW breaks valley degeneracy and results in an energy differential, $\Delta E^\tau$ of 0 - 20 meV between non-degenerate valleys. The strain energy is independent of the QW width unlike the quantum confinement energy which increases as the QW width decreases and is inversely proportional to the confinement mass; typical confinement energies $E_0^\tau$ are of the order of 20 meV for a narrow QW about 5 nm wide and 5 meV for a 20 nm wide QW. We defined a robust valley degeneracy condition to identify valleys which are degenerate independent of well width, and a crossover degeneracy condition where valleys are degenerate only for a particular crossover well width $W_0$. We showed that piezoelectric fields in the QW and barrier materials can arise in certain facets such as (111) and (411), and cause piezoelectric energy shifts $E_{\mathrm{pz}}^\tau$ as large as 50 meV for a (111) QW about 10 nm wide, and that such a piezoelectric field can be canceled by an external gate voltage. The smallest energy scale in these QWs is the Fermi energy which is a few meV depending on the degeneracy for electron densities around $n_{\mathrm{2D}} \sim 3 \times 10^{11} \, \mathrm{cm}^{-2}$ and assuming an effective mass $m_{\mathrm{2D}}^\tau \sim 0.3 m_e$. With such a small Fermi energy, valley occupation is strongly defined by the interplay of the various other energy scales in the problem. We once again emphasize that it is due to the similarity of all the energy scales that none of them can be ignored and must be treated carefully to calculate valley subband energy.

 Valley degenerate systems have an extra scattering channel not present in single valley systems, namely inter-valley scattering. We identified the main inter-valley scattering mechanisms as the interface roughness and alloy disorder scattering due to wavefunction penetration into the barrier and calculated the suppression factor at the crossover well width $W_0$ for inter-valley scattering between different sets of valleys. We also defined a valley scattering primitive unit cell to easily identify scattering events between the robust-degenerate and crossover-degenerate electron valleys. We drew the VSC for the special case of AlAs grown along four different orientations, the high-symmetry facets (001), (110), and (111), as well as a low-symmetry facet (411) to demonstrate the utility of our model. Furthermore, we explained the relevance of the VSC description for visual identification of anisotropic inter-valley scattering in AlAs QWs.

 In the final section we demonstrated the power of our formalism to determine valley degeneracy for arbitrary substrate orientation, without being restricted to Miller indexed planes as is prevalent in the literature. We consider the example of a miscut sample and define the additional notation required to address angle deviation from conventional growth axes. We then detail the procedure to calculate all the ground energy parameters as well as the strain tensor analysis for miscut samples.  It is worth repeating that the formalism developed here is sufficient to calculate valley degeneracy for {\em any} substrate orientation, and need not be aligned with any Miller index facet.

 There are some important limitations of our formalism. Firstly, we assume layer thicknesses are all within the strain relaxation limit such that there are no strain induced defects below or within the QW.  These would relax the lattice constant relative to the substrate, thus one would have to first determine the adjusted lattice constant at the QW layer and then apply the formalism developed here to deduce valley degeneracy. Secondly, we assume nearly empty QWs and do not calculate the Schr\"odinger equation self-consistently with the Poisson equation. However our formalism for calculating the various valley subband energies can be applied to standard self-consistent solvers to obtain the proper Hartree solution. Lastly, whereas the formalism takes into account the change in the confinement potential due to piezoelectric fields, it does not address larger structural considerations such as modulation doping layers, surface pinning potentials, electrostatic charges, and piezoelectric fields in the barrier, all of which would have to be calculated together to determine the confinement well potential and Fermi energy.  We note, however, that the piezoelectric equations provided here can be used to determine the piezoelectric fields in the strained barriers.

\section{Acknowledgments}\label{sec:Acknowledgments}
This work was funded by the NSF grant DMR-0748856, the MRSEC program of the National Science Foundation DMR-0520513 at the Materials Research Center of Northwestern University, and the BMBF Nano-Quit Project 01BM470. SPG and MG would like to thank Wade DeGottardi, Wang Zhou, Florian Herzog, and Jens Koch for helpful discussions.


\begin{thebibliography}{40}
\bibitem[{\citenamefont{GraysonAPLcomment}(1992)\citenamefont{GraysonAPLcomment}}]{GraysonAPLcomment}
 \bibinfo{journal}{Caridi and Stark incorrectly derived the equation for the strain tensor components of the (110) orientation.  Following the appendix of the present paper, the correct terms can be derived.
}
\bibitem[{\citenamefont{footnotebowing}}(1995)]{footnotebowing}
  \bibinfo{journal}{We have assumed that the bowing term from the expression for the bandgap in Ref. \cite{vurg} is accounted for entirely in the conduction band.}
\end{thebibliography}

\appendix*

\section{Analytical equations of the strain tensor and the strain ratios for arbitrary substrate orientations}


In Section \ref{sec:analyt}, we derive the strain ratios $D_i^M$ by minimizing the free energy of the strained layer. This section presents a generalized treatment to express these strain ratios for an arbitrary substrate orientation with cubic symmetry, given its coordinate transformation matrix ${\bf R}^M$, the elastic constants $c_{ijkl}$ in the CCC $\mathbf{x}$-basis and the in-plane strain $\epsilon'_{||}$ in the transport $\mathbf{a}$-basis.

We first obtain the fourth order rotated elastic stiffness tensor $c^{\prime}_{mnop}$ in the $\mathbf{a}$-basis by rotating the fourth-rank elastic stiffness tensor $c_{ijkl}$ from the CCC basis $\mathbf{x}=\left( x,y,z\right)$ to the transport basis $\mathbf{a}=\left( a,b,c\right)$
\begin{equation}
c^\prime_{mnop}=\sum_{ijkl}\mathrm{R}^M_{mi}\mathrm{R}^M_{nj}\mathrm{R}^M_{ok}\mathrm{R}^M_{pl}c_{ijkl}~~~,
\end{equation}
where $\mathrm{R}^M$ is the coordinate transformation matrix defined in Eq.~\eqref{equ:rotationmatrix}.
The rotated fourth-rank tensor $c^\prime_{mnop}$ in the CCC $\mathbf{x}$-basis should then be mapped to $C^{\prime}_{ij}$ using the Voigt notation.
The strain tensor components $\epsilon^{\prime} _{ij}$ in the transport $\mathbf{a}$-basis are given by
\begin{eqnarray}
\epsilon^{\prime} _{aa} &=&\epsilon^{\prime} _{bb}   =\epsilon^{\prime} _{\parallel} = \frac{a_{\rm{substrate}}-a_{\rm{layer}}}{a_{\rm{layer}}} \label{labelaa} \\
\epsilon^{\prime} _{ab} &=&\epsilon^{\prime} _{ba}   = 0 \label{labelab} \\
\epsilon^{\prime} _{ac} &=&\epsilon^{\prime} _{ca}   =\frac{\lambda \mu - \eta \omega}{\lambda \kappa - \eta^2}\epsilon^{\prime} _{\parallel} = D_1^M \epsilon^{\prime} _{\parallel}  \label{labelac} \\
\epsilon^{\prime} _{bc} &=&\epsilon^{\prime} _{cb}   =\frac{\omega - \eta D_1^M }{\lambda}\epsilon^{\prime} _{\parallel} = D_2^M \epsilon^{\prime} _{\parallel}   \label{labelbc} \\
\epsilon^{\prime} _{cc} &=&\epsilon^{\prime}_{\perp} =\frac{\alpha-2C_{34}^{\prime}D_2^M-2C_{35}^{\prime }D_1^M}{C_{33}^{\prime}} \epsilon^{\prime} _{\parallel}\\
&=& -D_0^M \epsilon^{\prime} _{\parallel} \notag \label{labelcc}
\end{eqnarray}
where the denominators in Eq.~\eqref{labelac}, Eq.~\eqref{labelbc} and  Eq.~\eqref{labelcc} are always nonzero, the coefficients $\alpha$, $\beta$ and $\gamma$ are first order, and the coefficients $\lambda$, $\kappa$, $\eta$, $\omega$ and $\mu$ are second order in the elastic stiffness tensor components $C_{ij}^{\prime}$
\begin{eqnarray}
\alpha  &=& - \left( C^{\prime}_{13} + C^{\prime}_{23} \right) \\
\beta   &=& - \left( C^{\prime}_{14} + C^{\prime}_{24} \right) \\
\gamma  &=& - \left( C^{\prime}_{15} + C^{\prime}_{25} \right) \\
\lambda &=& 2        C^{\prime}_{33}C^{\prime}_{44} \\
\kappa  &=& 2        C^{\prime}_{33}C^{\prime}_{55} \\
\eta    &=& 2\left( C^{\prime}_{33}C^{\prime}_{45} - C^{\prime}_{34}C^{\prime}_{35} \right) \\
\omega  &=& C^{\prime}_{33}\beta  - C^{\prime}_{34}\alpha \\
\mu     &=& C^{\prime}_{33}\gamma - C^{\prime}_{35}\alpha ~~~.
\end{eqnarray}%

\end{document}